\documentclass[12pt,preprint]{aastex}

\shorttitle{Testing the GRB Pulse Start Conjecture}
\shortauthors{Hakkila et al.}

\begin{document}

\title{Testing the Gamma-Ray Burst Pulse Start Conjecture}

\author{Jon Hakkila\altaffilmark{1} and Robert J. Nemiroff\altaffilmark{2}} 

\affil{$^1$Dept. Physics and Astronomy, The College of Charleston, Charleston, SC  29424-0001\\$^2$Dept. Physics and Astronomy, Michigan Technological University, Houghton, MI \\}
\email{hakkilaj@cofc.edu}

\begin{abstract}

We test the hypothesis that prompt gamma-ray burst pulse emission 
starts simultaneously at all energies (the Pulse Start Conjecture). Our analysis,
using a sample of BATSE bursts observed with four channel, 64-ms data and 
performed using a pulse fit model, generally supports this hypothesis for the Long GRB class, although
a few discrepant pulses belong to bursts observed during times
characterized by low signal-to-noise, hidden pulses, and/or significant pulse overlap. The typical
uncertainty in making this statement is $ < 0.4$ s for pulses in Long GRBs (and 
$< 0.2$ s for $40\%$ of the pulses) and
perhaps $ < 0.1$ s for pulses in Short GRBs.
When considered along with the $E_{\rm pk}$ decline
found in GRB pulse evolution, this result implies that energy is injected at 
the beginning of each and every GRB pulse, and the subsequent spectral evolution,
including the pulse peak intensity, represents radiated energy losses from
this initial injection.

\end{abstract}




\section{Introduction}

The physical mechanisms that create prompt high energy emission in gamma-ray bursts (GRBs) remain unknown even 40 years after their discovery.  The most commonly discussed models typically involve colliding shocks in jets of relativistically moving material (for a review, see \cite{pir05}).  It has been known for quite some time, however, that GRB light curves usually contain any number of similar sub-structures known as {\em pulses} \citep{des81}.  \cite{nor96} closely inspected 41 GRBs detected by the Burst and Transient Source Experiment (BATSE) onboard the Compton Gamma Ray Observatory and decomposed them into over 400 constituent pulses, correlating their properties.  

The simple nature of pulses and their ubiquitous appearance in both the Long and Short GRB classes (e.g. \cite{kou93}) have led to many studies focusing on them.  Pulses from the Long GRB class have been studied in BATSE data much more than pulses from Short GRBs due to the limited amount of high time resolution BATSE data, and to the low-energy sensitivity of recent experiments such as Swift and HETE-2.

Many pulses from Long GRBs appear to be a good statistical fit to a simple temporal form \citep{nor05, hak08, hak09}, have shorter durations at higher energies \citep{nor05, ryd05a}, appear time asymmetric \citep{nem94, nor02, koc03, nor05}, and become softer as they progress \citep{ryd05a, hak09}. For most pulses these correlated, energy-integrated properties are supportive
of the hypothesis that pulses are scalable versions of one another \citep{nor96, ste96, nem00, koc03}. 
Exceptions may be faint and/or very short events and peaked, long duration pulses \citep{hak08, hak09}. 
Pulse spectra {\em evolve} with time in a systematic fashion:
a pulse peaks first at high energies and later at lower energies; 
this can be measured from pulse peak lags \citep{hak08, hak09} or by the decay of $E_{\rm pk}$ (the peak of the $\nu F_{\nu}$ spectrum) (e.g. \cite{kan06, pen09}).

Faint pulses have longer durations than bright pulses \citep{rrm00, hak09}; this is because pulses with short spectral lags and durations are more luminous than those with long spectral lags and durations \citep{hak08}. Since pulse properties of asymmetry and spectral hardness anti-correlate with duration, these are also luminosity indicators \citep{hak09}. In particular, the statistical lag between the bulk of GRB emission received at low energies relative to high energies has been used as an indicator of intrinsic GRB luminosity \citep{nor00}, an effort that has recently been redirected on the statistical lags of constituent pulses \citep{hak09b}.   Such efforts make GRBs into a standardizable candle that can be used to estimate the composition and geometry of the universe at distances further than supernovae (see, for example, \cite{sch07}).

Simple correlations among pulse characteristics may explain many seemingly complex behaviors of bulk GRB prompt emission. For example, the relationship between burst lag vs.\  peak luminosity \citep{nor00} is fundamentally a property of pulses.  Pulses within a burst can have diverse lags, and the bulk lag bears a complex relationship to the pulse lags \citep{hak08}.  Since pulse lag is directly related to pulse duration and since both vary inversely with pulse peak luminosity, the bulk lag for a burst, obtained from the cross-correlation function \citep{ban97}, tends to be biased towards the highest intensity, shortest pulses, while the peak luminosity of a burst measures the overlapping intensities of disparate pulses.  The bulk lag vs.\ luminosity relation is thus a manifestation of a more fundamental pulse relation, but distorted in a complex way.  Similarly, the the variability vs.\ luminosity relation \citep{rei01} is actually a measure of pulse properties since variability is related to the number of pulses and durations of the pulses in a burst, and the $E_{\rm pk}$ vs. $E_{\rm iso}$ relation \citep{ama02} is a measure of pulse values since $E_{\rm pk}$ is a time-integrated value constructed from the merged spectra of many pulses. As bursts are linear combinations of pulses, bulk properties represent a kind of average that destroys information, suggesting that pulse-level studies identify stronger and/or tighter correlations than those relying on bulk properties. 

In principle, the bulk characteristics of the prompt emission can be derived from knowledge of pulse properties and the pulse decomposition of a burst. The converse is not true: we cannot infer the basic characteristics of the pulses, nor can we understand the relationship between bulk and (more fundamental) pulse properties, without direct, explicit study of the constituent pulses.

Characterization of pulse attributes help define and constrain physical models.  \cite{kat94} and others made early suggestions that GRB pulse shapes originate from time delays inherent in the geometry of spherically expanding emission fronts. \cite{lia97} provided arguments that saturated Compton up-scattering of softer photons may be the dominant physical mechanism that creates the shape of GRB pulses. The similar time evolution of pulse structures, combined with the fact that 
their measurable properties correlate strongly, suggests that one physical 
mechanism produces the array of pulse characteristics. There is strong evidence 
that the majority of GRB pulses result from internal shocks in relativistic winds; 
these arguments have been made on the short durations, spectral evolution, 
and short interpulse durations (e.g. \cite{dai98, rrm00, nak02}). \cite{sum97} and \cite{rrm00} claim that the rise time of GRB fast rise exponential decay (FRED)-like structures is related to the sound speed of the pulse medium, but the decay time is related to a time delay inherent in the geometry of an expanding, spherical GRB wave front undergoing rapid synchrotron cooling. 

Pulse evolution is clearly important to understanding the physics of GRB pulses (and thus of GRB prompt emission). Although many questions have been answered, one important question about the onset of GRB pulses remains.  \cite{nem00} proposed the Pulse Start Conjecture; namely, that a single GRB pulse starts simultaneously at all energies.  The semi-automated pulse fitting procedure of \citep{hak08} provides us with a mechanism for testing the Pulse Start Conjecture, and the beginning pulse database created to date from application of this model to multi-channel 64 ms BATSE data \citep{hak09} provides a rudimentary database for testing this conjecture.

\section{Analysis}

\subsection{Pulse Fitting Methodology}

Our pulse-fitting technique begins with a search for pulses in summed four channel BATSE data. Candidate time intervals potentially containing pulses are identified using the Bayesian Blocks methodology \citep{sca98}; suspected pulses within each interval are modeled using the four-parameter pulse model of \cite{nor05} and fitted using the iterative Levenberg-Marquardt algorithm found in the IDL subroutine MPFIT \citep{mar08}. 
The \cite{nor05} pulse model assumes each pulse can be fit by:
\begin{equation} 
I(t) = A \lambda \exp^{[-\tau_1/(t - t_s) - (t - t_s)/\tau_2]},
\end{equation}
where $t$ is time since the trigger, $A$ is the pulse amplitude, $t_s$ is the pulse start time, 
$\tau_1$ and $\tau_2$ are characteristics of the pulse rise and pulse decay, and 
$\lambda = \exp{[2 (\tau_1/\tau_2)^{1/2}]}$). We use a two-parameter model to fit the background
along with the pulses.

Prior to searching for pulses, Poisson noise is added to time intervals having 1 s time resolution. Intervals having these resolutions occur more than 2 s before the trigger, and sometimes late in the evolution of very long bursts (later than roughly 90 minutes after the trigger). This 'noisification' allows all time intervals to be treated in a statistically-similar fashion by the fitting routines.

During the pulse identification phase, some of the Bayesian Block intervals are found to contain statistically-insignificant modeled pulses that are removed from consideration using a threshold based on two timescales \citep{hak03}. This {\em dual timescale} threshold does not favor short duration, high-amplitude pulses or long duration, low-amplitude pulses over one another. After each iteration in which insignificant pulses are removed, the fitting process begins again. Iteration eventually produces an optimal set of fitted pulses for the summed four-channel data. 

The process of finding pulses in individual energy channels is initiated with the pulse parameters obtained from the summed four-channel data; the identified pulses are scaled to the single-channel count rates and this is the first guess as to the pulse characteristics. Iteration proceeds as described previously, with insignificant pulses being removed. After iterating on all four individual energy channels, a final set of fits is obtained for each pulse in each energy channel in each GRB. 

There is generally little ambiguity in establishing that pulses across different energy channels are the same pulse. The fitting process, combined with the initial pulse parameter guesses, assumes that each pulse peaks at identical times in all energy channels, and the Levenberg-Marquardt iteration process does not cause subsequent guesses to stray far enough from the initial ones to confuse the peaks of unrelated pulses. High signal to noise ratios generally found in BATSE channels 1 through 3 and sometimes found in channel 4 also prevent this confusion. There is no ambiguity when fitting pulses that are isolated or which overlap little with other pulses; the greatest ambiguity occurs for pulses that are faint in one or more energy channels, and/or which overlap significantly in time with other pulses. To prevent incorrect pulse identification, we exclude from our database pulses that we cannot clearly associate across a range of energy channels due to faintness or pulse overlap.

As mentioned above, we add Poisson noise for times more than 2 seconds before the trigger. This noise changes the background in a random way, but for the most part retains the underlying pulse structure that can be recognized by the pulse-fitting routine. We have found that the addition of this noise has essentially no impact on pulses beginning after $t=-2$ s, and generally has minor impact otherwise. Poisson variations in the background  provide less of a spurious signal than, for example, the contribution from an overlapping pulse or from a non-Poisson source (such as Vela X-1, Cygnus X-1, or an occultation step). Monte Carlo runs demonstrate that the start time typically moves only a small amount. Thus, we feel justified in neglecting the results of this effect.

The pulse extraction procedure described here is fairly successful, even when the pulse signal-to-noise ratio is low in more than one of the BATSE energy channels (which often happens in channel 4). Pulse properties are cleanly extracted in a moderately unambiguous manner for many GRBs containing non-overlapping or isolated pulses. Typically, these ``clean'' pulses belong to low luminosity, long duration bursts \citep{nor05, hak07}. However, the process is also successful at identifying and fitting many pulses in complex GRBs containing overlapping pulses. The approach is less successful at fitting low intensity pulses, pulses that ambiguously overlap, and  very short pulses, which can be indistinguishable from Poisson noise. Ambiguous pulse identifications often result in poor $\chi^2$ goodness-of-fit measures, in fits that appear to merge pulses separable to the eye, and/or in pulses that have disparate properties or are not observed in contiguous energy channels. We exclude from our analysis overlapping and low fluence pulses that are overtly ambiguous and that cannot be clearly identified in consecutive energy channels.

We have used machine learning algorithms \citep{hak03, hak07} to further delineate BATSE GRBs into two classes based on duration and spectral hardness. A GRB belongs to the Short class if it satisfies the inequality  (T90 $<1.954$) OR ($1.954 \le T90 < 4.672$ AND ${\rm HR}_{321} > 3.01$) (where ${\rm HR}_{321} = S_3/(S_2 + S_1)$ \citep{muk98}; $S_i$ is the channel $i$ fluence); otherwise it belongs to the Long class. This classification scheme inherently assumes that the Intermediate GRB class (e.g. \cite{hor98, muk98, hak00}) is caused by sampling biases related to a peak flux trigger combined with a fluence classification parameter. It also oversimplifies the region where the Long and Short GRB classes overlap by replacing an inherently fuzzy boundary with a somewhat arbitrary sharp one. Nonetheless, it allows us to delineate Long and Short GRBs in our sample.

Our prior analyses using this technique have indicated that pulse properties, rather than bulk properties of the prompt emission, underly GRB measurements \citep{hak08}. Most pulses belonging to the Long class of GRBs are found to have highly correlative properties (exceptions are some long duration pulses with high-intensity, short-lag peaks) while the correlative properties of Short GRB pulses may be weaker but are also harder to determine due to BATSE's temporal resolution \citep{hak09}.

\subsection{Pulse Start Times}

\cite{nor05} recognized the need for pulse start time $t_s$ as a fitting parameter in the pulse model (Eqn. 1), yet calculated several pulse observables (e.g. pulse duration, pulse peak flux, pulse asymmetry) without explicitly mentioning the contribution of $t_s$. They seem to have done this because the pulse start time seems almost of secondary importance to the process of fitting the pulse: it defines the time that the flux statistically rises above background and is mainly needed because the fitting function is undefined and blows up for $t < t_s$. As a result, \cite{nor05} measured the uncertainties in these observables without formally propagating any pulse start time uncertainties. All pulse-fitting errors and the derived observables were assumed to come from Gaussian error distributions, and these distributions proved effective at measuring and characterizing pulse properties for the long duration pulses studied in their sample. Due to the \cite{nor05} success at using this approach, our technique for fitting GRB pulses has also assumed Gaussian error distributions for all pulse properties, including those of the previously excluded pulse start times. 

Our first hint that the assumption of Gaussian error distributions might not always be completely valid occurs when measuring the energy-dependent pulse peak times (the basis for measuring pulse peak lags). The sometimes large formal uncertainties in the pulse start times $t_{s;e}$ for each each energy channel $e$ lead to inordinately large formal uncertainties in the calculation of the pulse peak times, so Monte Carlo error analysis has been used instead (e.g. \citep{hak07, hak09}). In testing the Pulse Start Conjecture, we have chosen to use the formal Gaussian pulse start time uncertainties returned by MPFIT in order to better understand any systematic biases in the measurement of $t_{s;e}$.

Table \ref{tbl-1} contains start times for 199 pulses in 75 Long BATSE GRBs, and Table \ref{tbl-2} contains start times for 41 pulses in 33 Short BATSE GRBs, along with their formal uncertainties. Although start times have been measured for most of these pulses, many Long GRB pulses do not have sufficient channel 4 flux for pulse fitting, thus limiting the usefulness of pulse start time measurement at these energies. Our sample has been obtained by attempting to fit GRB pulses sequentially through the BATSE Catalog \citep{pac99}, in order to minimize sampling biases. We have excluded GRBs only if we could not obtain a believable fit; in other words, if all fit pulses seemed ambiguous. We note that a very large range in pulse start time uncertainties exists. We comment on this below.

\subsection{Testing Consistency Between Energy-Dependent Pulse Start Times}

Our goal is to determine if the pulse start times measured in different energy channels are consistent with a single pulse start time. This is akin to asking if the uncertainty in start times measured for a pulse across each energy channel is less than or equal to that predicted from the formal uncertainty of this measurement.  Thus, we use both the multi-wavelength pulse start times and the uncertainties in these start times to test our hypothesis.

For each individual pulse, the start time in energy channel $e$ measured from the pulse fit model is given as $t_{s; e}$. The pulse-fitting model, through the subroutine MPFIT, calculates a formal uncertainty of $\sigma_{t_{s ; e}}$ for this measured value, as it calculates formal uncertainties for all pulse parameters and for the background.

A value of $\chi_i^2$ per degree of freedom can be calculated across each of $n$ energy channels in a particular pulse $i$ by 
\begin{equation}
\chi_i^2 = \frac{1}{n-1}\sum_{e=1}^n (\frac{t_{s ; e}-\bar{t}_s}{\sigma_e})^2,
\end{equation}
where we test the hypothesis by assuming that a single start time fits the multi-channel data and that this mean start time $\bar{t}_s$ is obtained by averaging the start times obtained from the individual energy channels.

We assume that each energy channel contains independent information pertaining to the pulses. This is a good first-order approximation because the BATSE detector response matrices are roughly diagonal, although high energy photons can have their energy deposited in lower-energy BATSE channels \citep{pen99}. Thus, we assume that the pulse start time uncertainties in different energy channels are independent of one another.

\subsubsection{Long GRB Pulses}

\begin{figure}
\plottwo{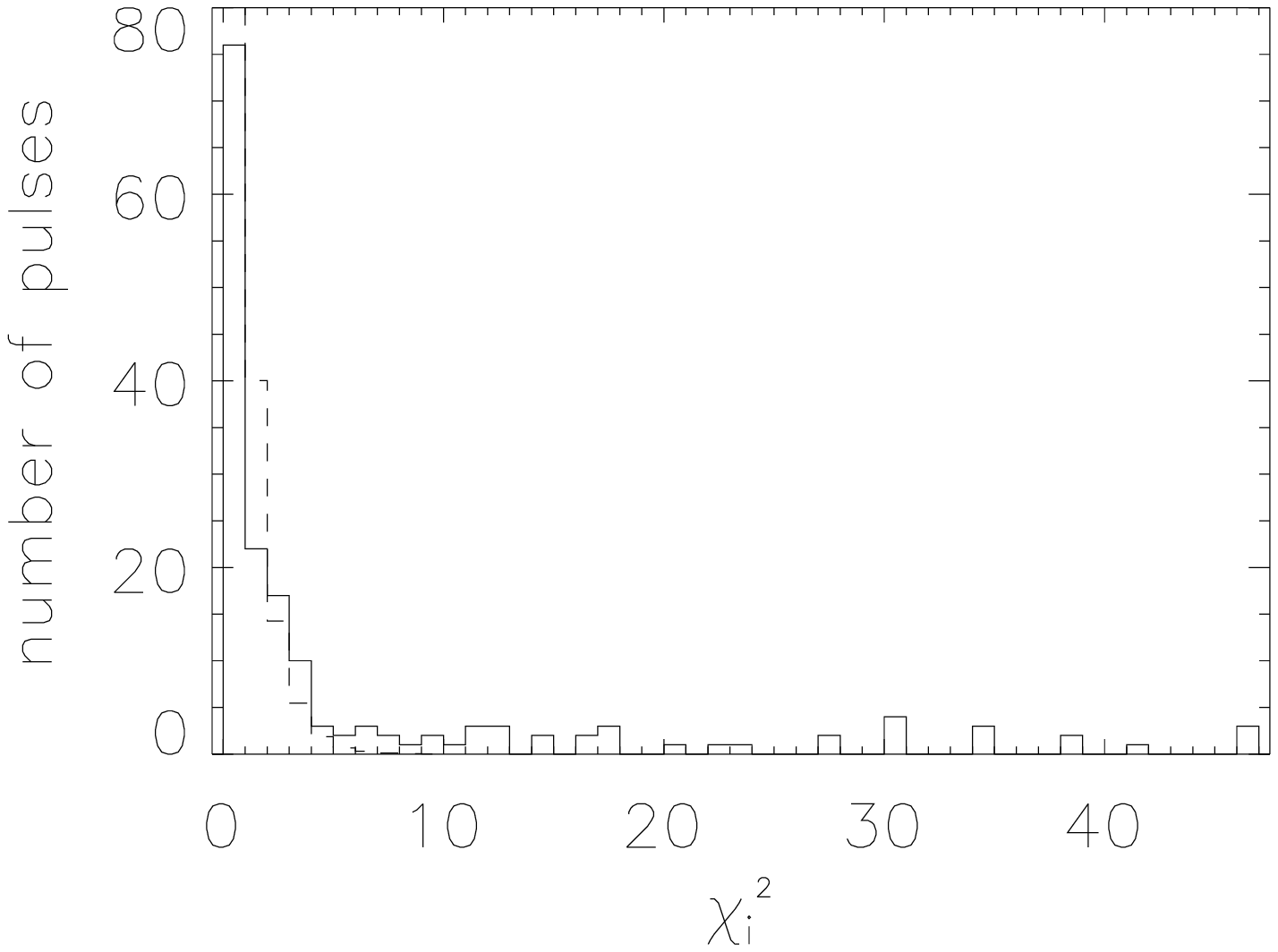}{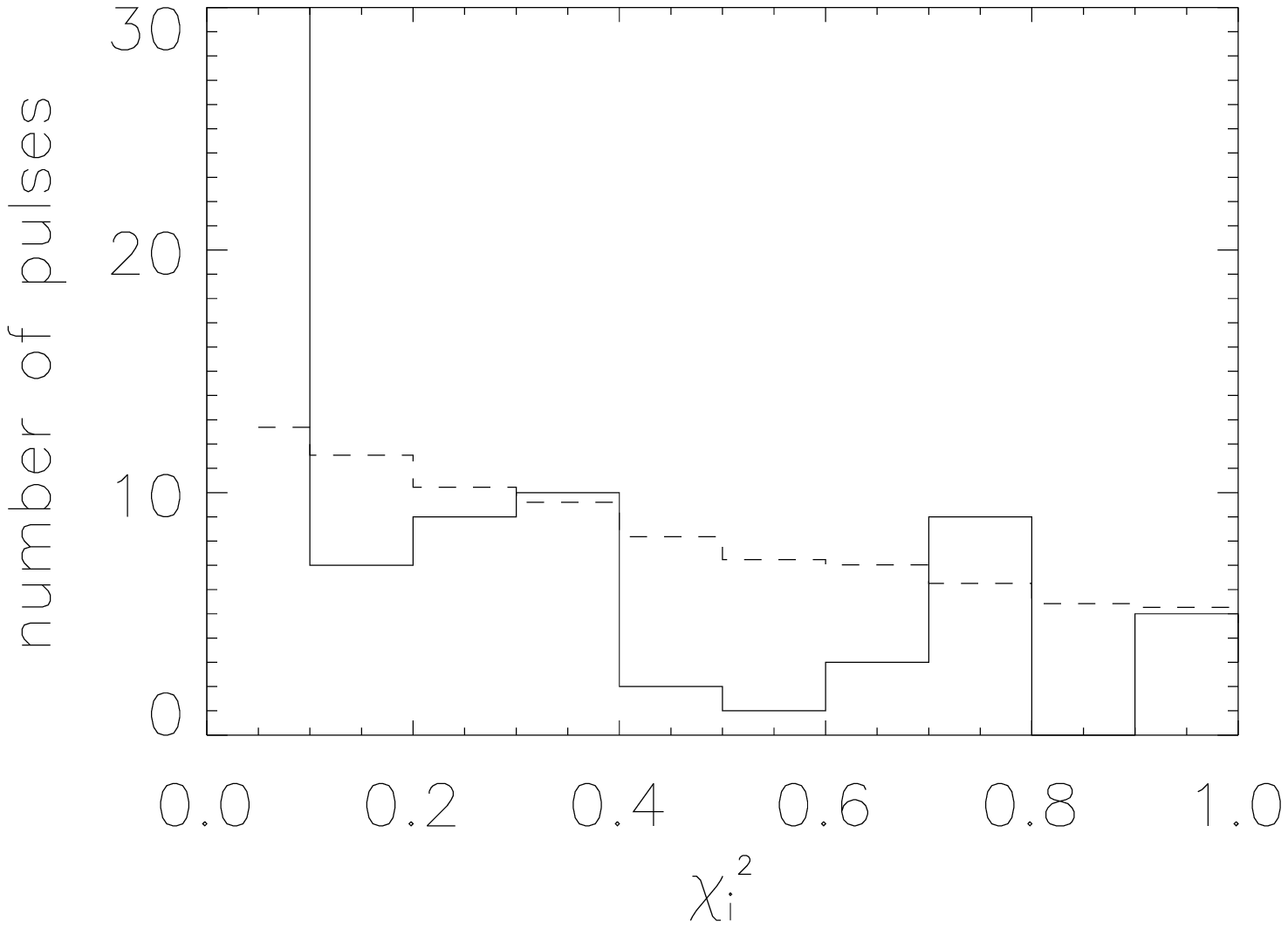}
\caption{The distribution of $\chi_i^2$ (solid histogram) for combined Long GRB pulse start times across energy channels 1, 2, and 3 (equation 2) is compared to the expected distribution (dotted histogram). There are excesses of pulses having $\chi_i^2$ values that are both too large (left panel) and too small (right panel) - many of the largest $\chi_i^2$ values are not plotted because they extend beyond the maximum plotted value. \label{fig1}}
\end{figure}

The $\chi_i^2$ distribution for the sample of Long GRB pulses described in Table \ref{tbl-1} is shown in Figure \ref{fig1}.  Channel 1, 2, and 3 pulse start times have been used in this analysis; channel 4 start times have been excluded because a) few Long GRB  pulses have been fitted in channel 4, and b) channel 4 start time uncertainties are often abnormally large for the fitted pulses. 

The behavior of the $\chi_i^2$ distribution (solid histogram) deviates from the theoretical distribution (dashed histogram) because there are too many pulses having both unexpectedly large $\chi_i^2$ values (Figure \ref{fig1}a) and unexpectedly small ones (Figure \ref{fig1}b). Large $\chi_i^2$ values indicate pulses that start at statistically different times in each energy channel. Too many pulses with large $\chi_i^2$ values could indicate that the Pulse Start Conjecture is false, but these pulses could also indicate other systematic difficulties in fitting pulses, such as low signal-to-noise, pulse overlap, pulses observed simultaneously with another active background sources, or other pulse fitting problems. Very small $\chi_i^2$ values identify either pulses that clearly satisfy the Pulse Start Conjecture, or pulses whose pulse start time uncertainties have been systematically overestimated by the fitting process; these pulses appear to start simultaneously in all energy channels. The $\chi_i^2$ distribution needs to be understood before the Pulse Start Conjecture can be accurately assessed.

Pulses having the largest or smallest start time uncertainties have correspondingly extreme values of the pulse rise variable $\tau_1$  ($\tau_1>10^4$ or $\tau_1< 10^{-5}$).  That $t_s$ and $\tau_1$ are coupled is not surprising since each of the four pulse fitting variables primarily describes a different temporal pulse region: a) $t_s$ divides the time before the pulse from the time during the pulse, b) $\tau_1$ describes the pulse during the time of the pulse rise, c) the amplitude $A$ is indicated at the time of pulse peak, and d) $\tau_2$ describes the pulse during the time of the pulse decay. Thus, $t_s$ and $\tau_1$ are closely related because both fit the pulse prior to the pulse peak. A rapid pulse rise suggests that that pulse start time is well-defined, while a gradual pulse rise rate suggests that the pulse start time is poorly constrained. The rapidly-varying pulse functional form (Equation 1) coupled with the minimum 64 ms temporal resolution, low signal to noise ratios, and double precision limitations of our fulse-fitting code, can combine to generate very long or very short pulse rise times, extreme $\tau_1$ values, and either under-constrained or over-constrained pulse start time uncertainties. As a result, pulses with very large or very small start time uncertainties have limited usefulness in constraining the Pulse Start Conjecture, and should be excluded from the analysis. 

In order to avoid the aforementioned concerns, we have avoided pulses having very large or very small pulse start time uncertainties in one or more energy channels. Our data paring is considered complete when the resulting $\chi_i^2$ distribution appears reasonably similar to a theoretical one. We find that our fit becomes reasonable when we exclude pulses having: a) $\sigma_e > 6$ s in at least one energy channel (an uncertainty often longer than the duration of the pulse), b) $\sigma_e < 0.032$ s in at least one energy channel (corresponding to temporal resolution of half a time bin), and c) $\chi_i^2 > 5$ values that can otherwise be attributed to systematic effects. This last condition requires us to closely examine fits of the pulses that potentially do not satisfy the Pulse Start Conjecture; proof or disproof of the Conjecture could potentially rest on proper interpretation of these fits. 

\begin{figure}
\plottwo{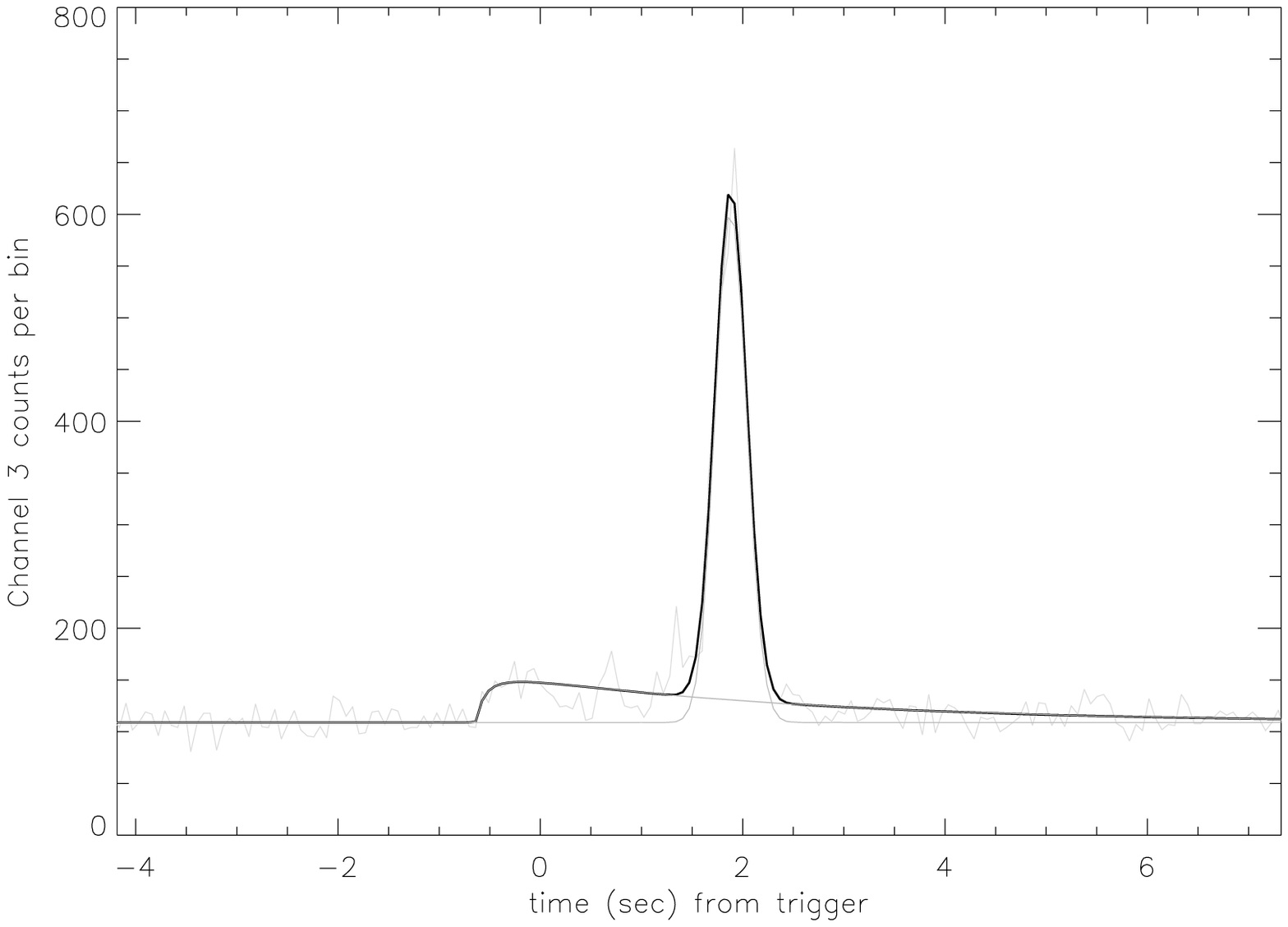}{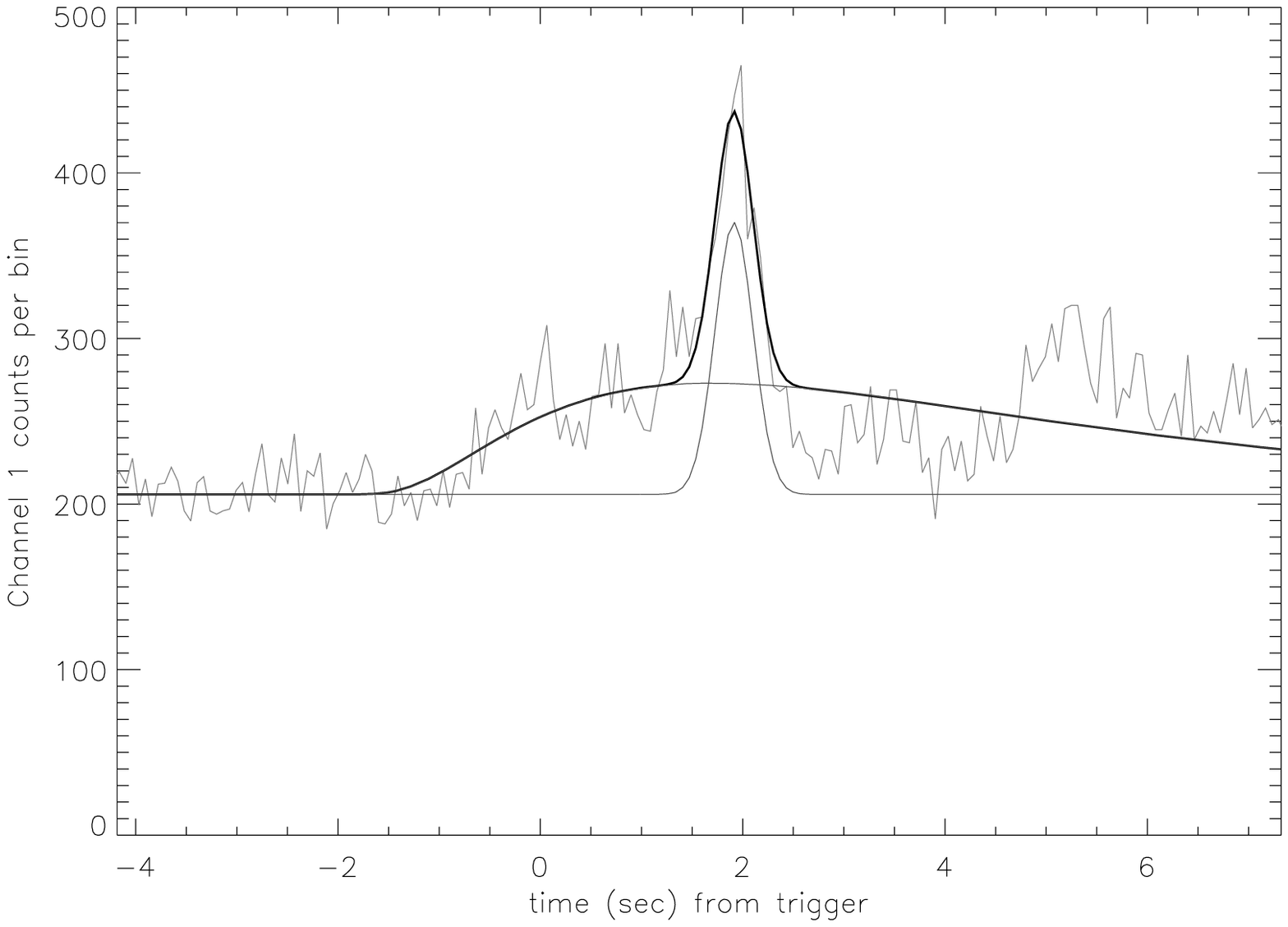}
\caption{Discordant pulse start times in channel 3 (left panel) and channel 1 (right panel) for pulse 1 of BATSE trigger 179.  Pulse 1 is the low amplitude pulse that triggered BATSE. Table \ref{tbl-3} explains why the pulse is subsequently removed from the sample. \label{fig2}}
\end{figure}

\begin{figure}
\plottwo{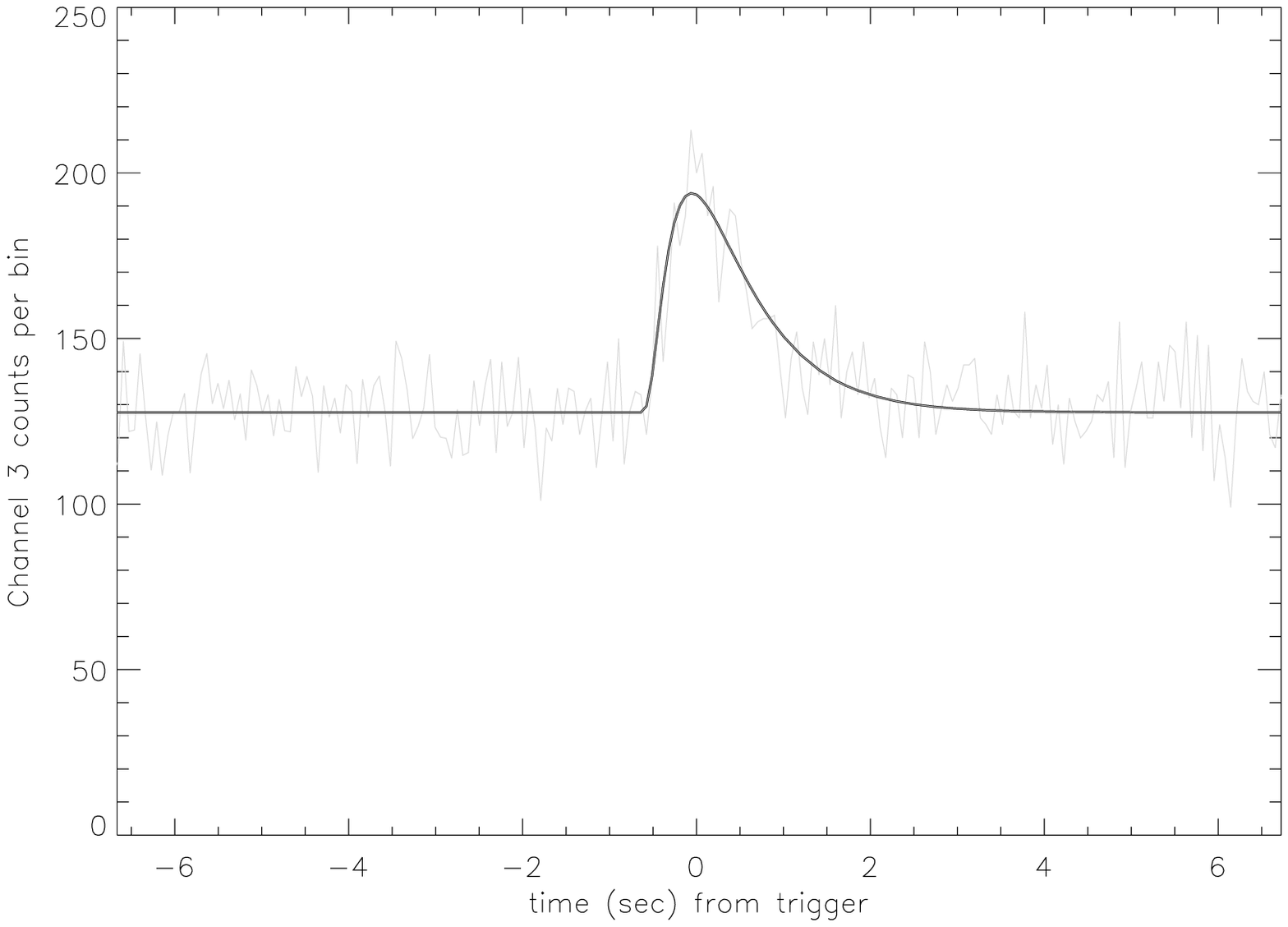}{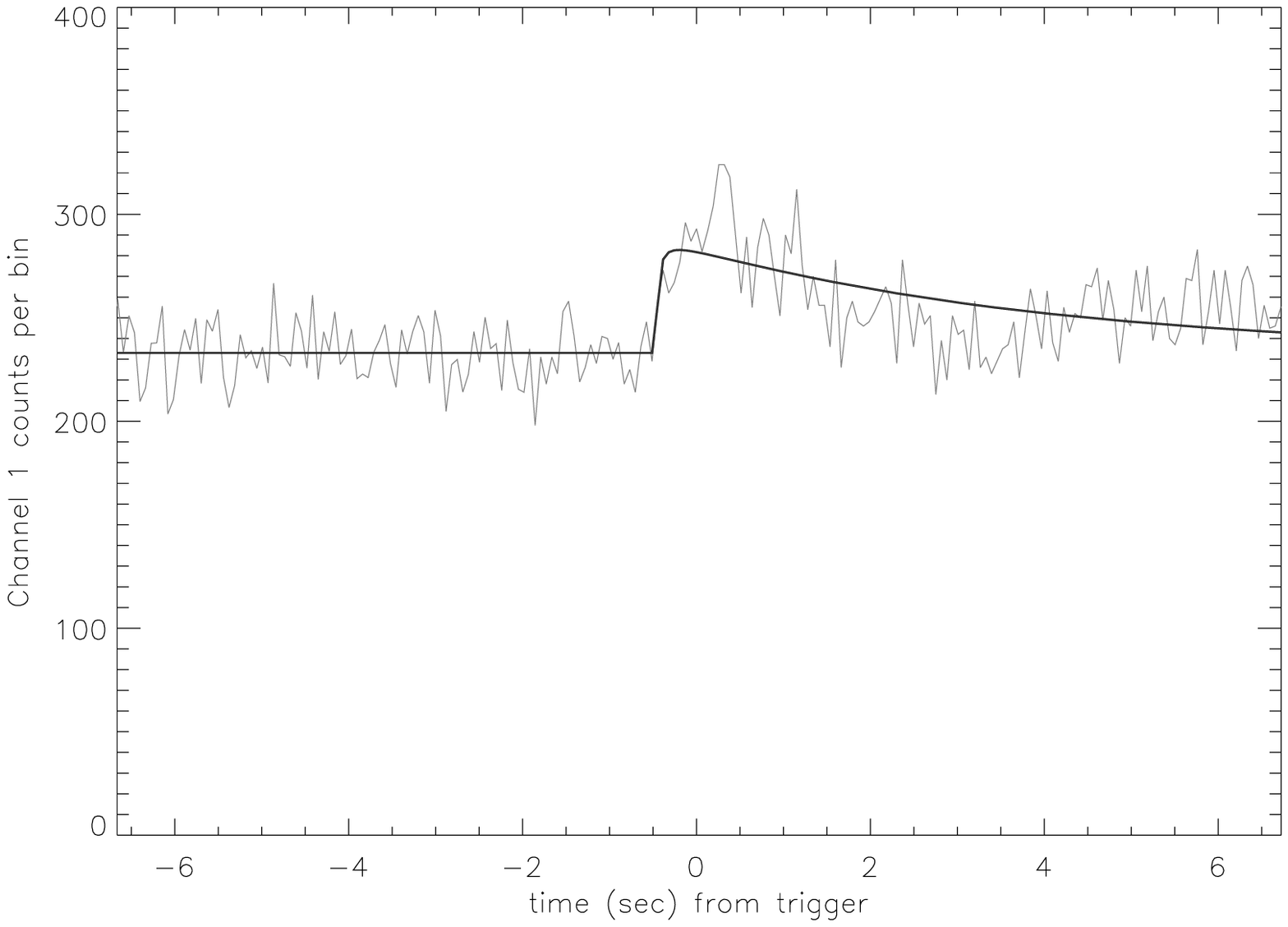}
\caption{Discordant pulse start times in channel 3 (left panel) and channel 1 (right panel) of BATSE trigger 228. Table \ref{tbl-3} explains why the pulse is subsequently removed from the sample.\label{fig3}}
\end{figure}

\begin{figure}
\plottwo{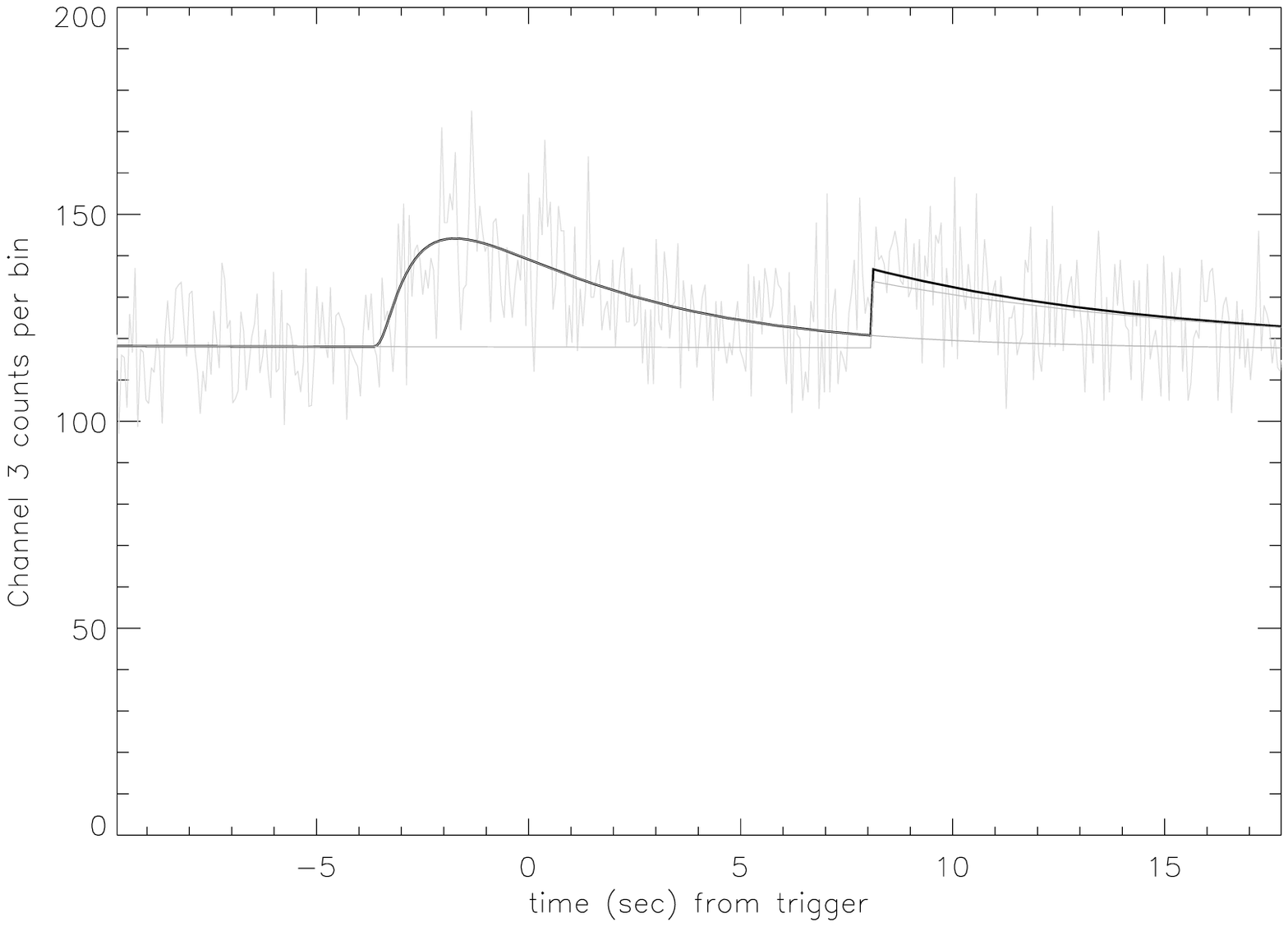}{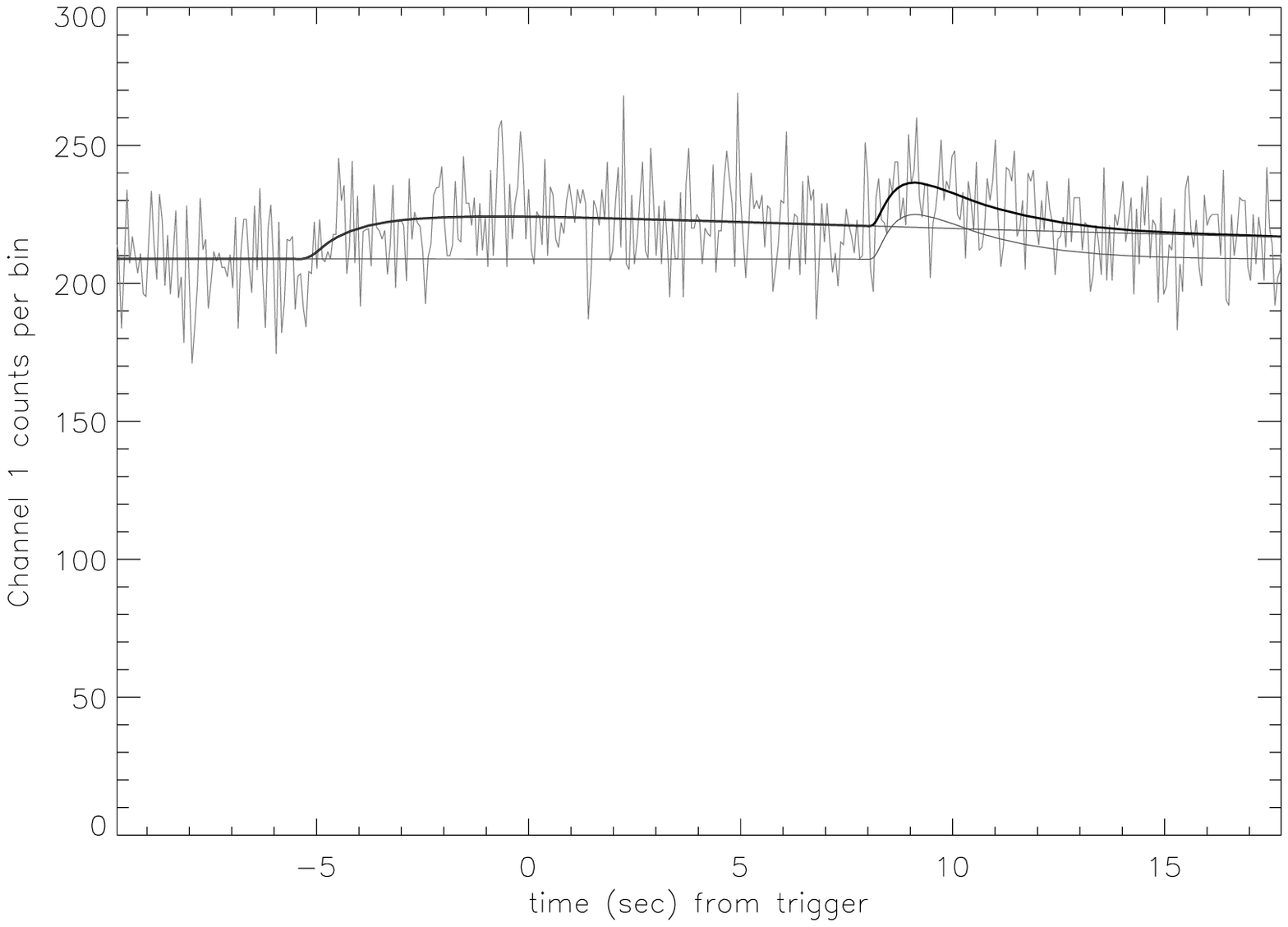}
\caption{Discordant pulse start times in channel 3 (left panel) and channel 1 (right panel) for pulse 1 of BATSE trigger 288.  Pulse 1 is the hard initial pulse that caused the BATSE trigger. Table \ref{tbl-3} explains why the pulse is subsequently removed from the sample.\label{fig4}}
\end{figure}

\begin{figure}
\plottwo{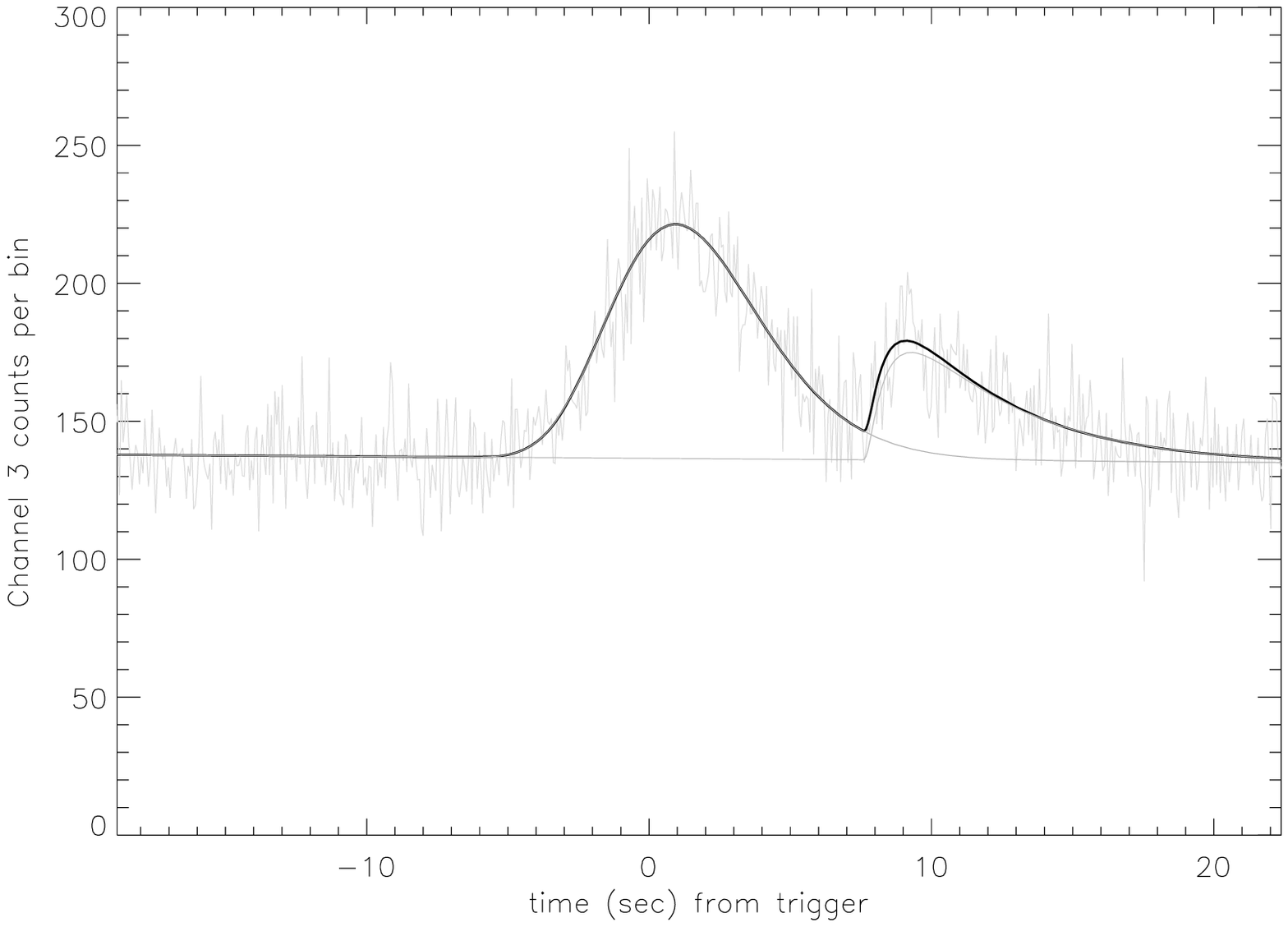}{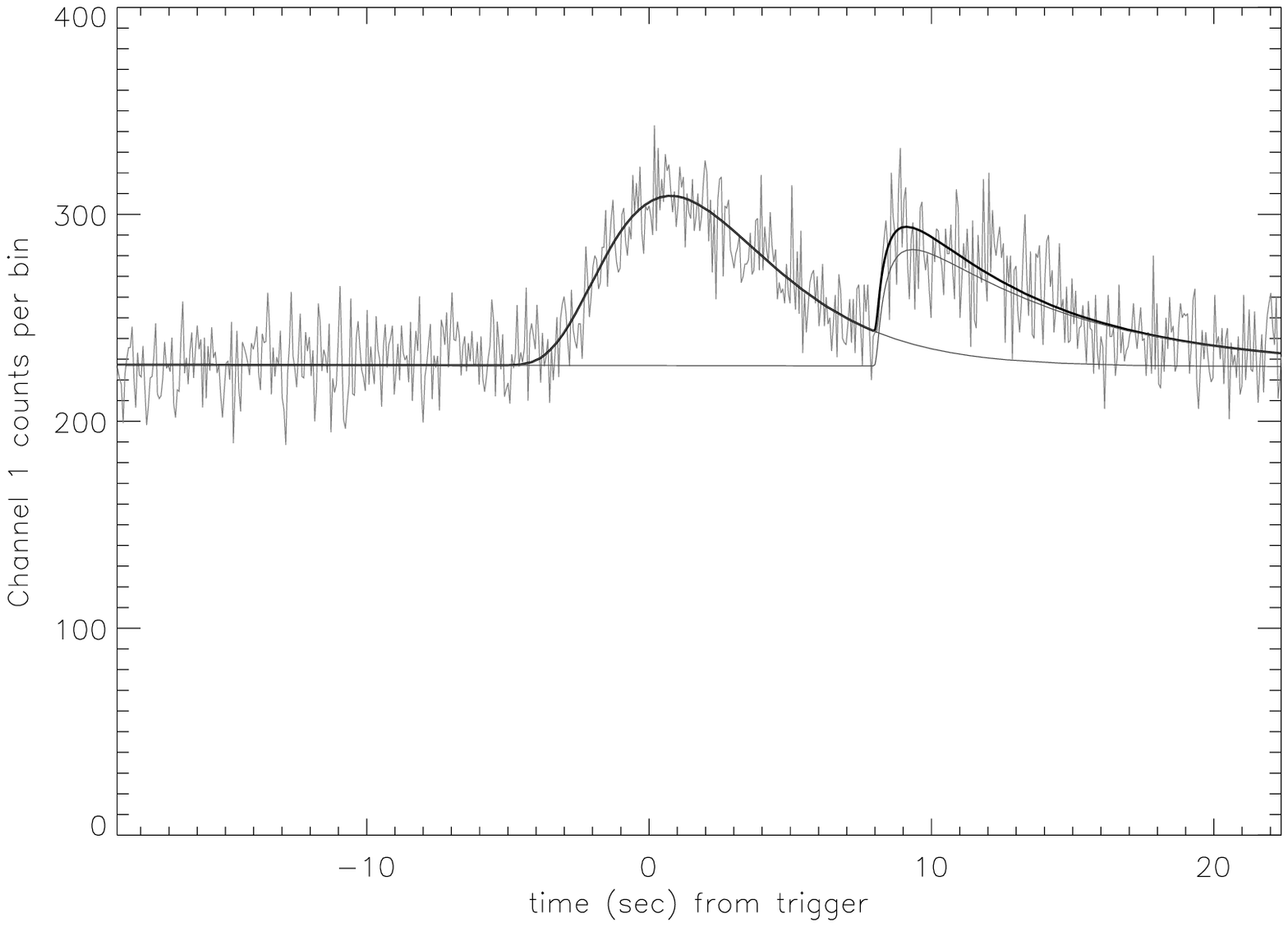}
\caption{Discordant pulse start times in channel 3 (left panel) and channel 1 (right panel) for pulse 2 of BATSE trigger 398.  Pulse 2 is the low amplitude secondary pulse. Table \ref{tbl-3} explains why the pulse is subsequently removed from the sample.\label{fig5}}
\end{figure}

\begin{figure}
\plottwo{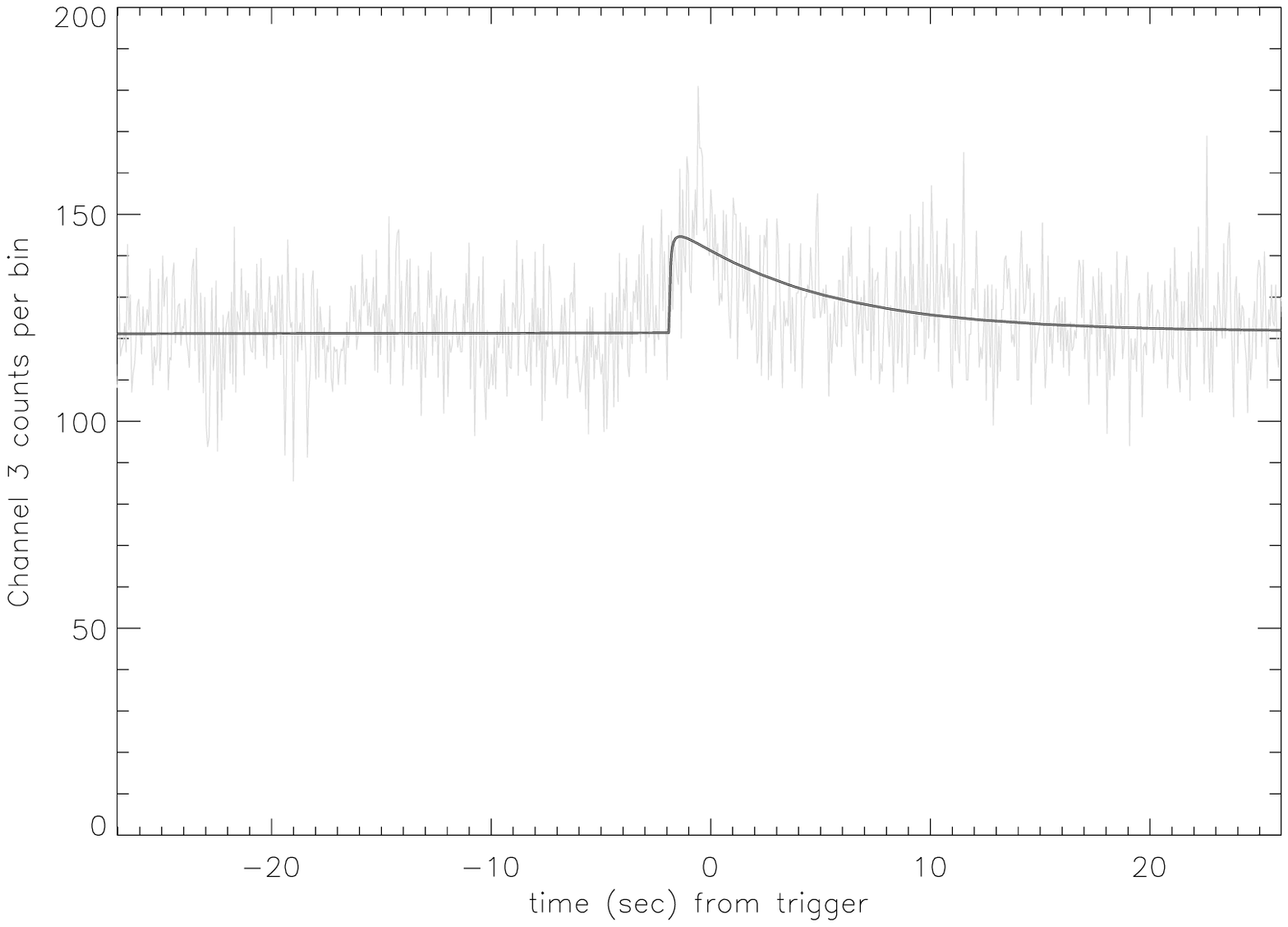}{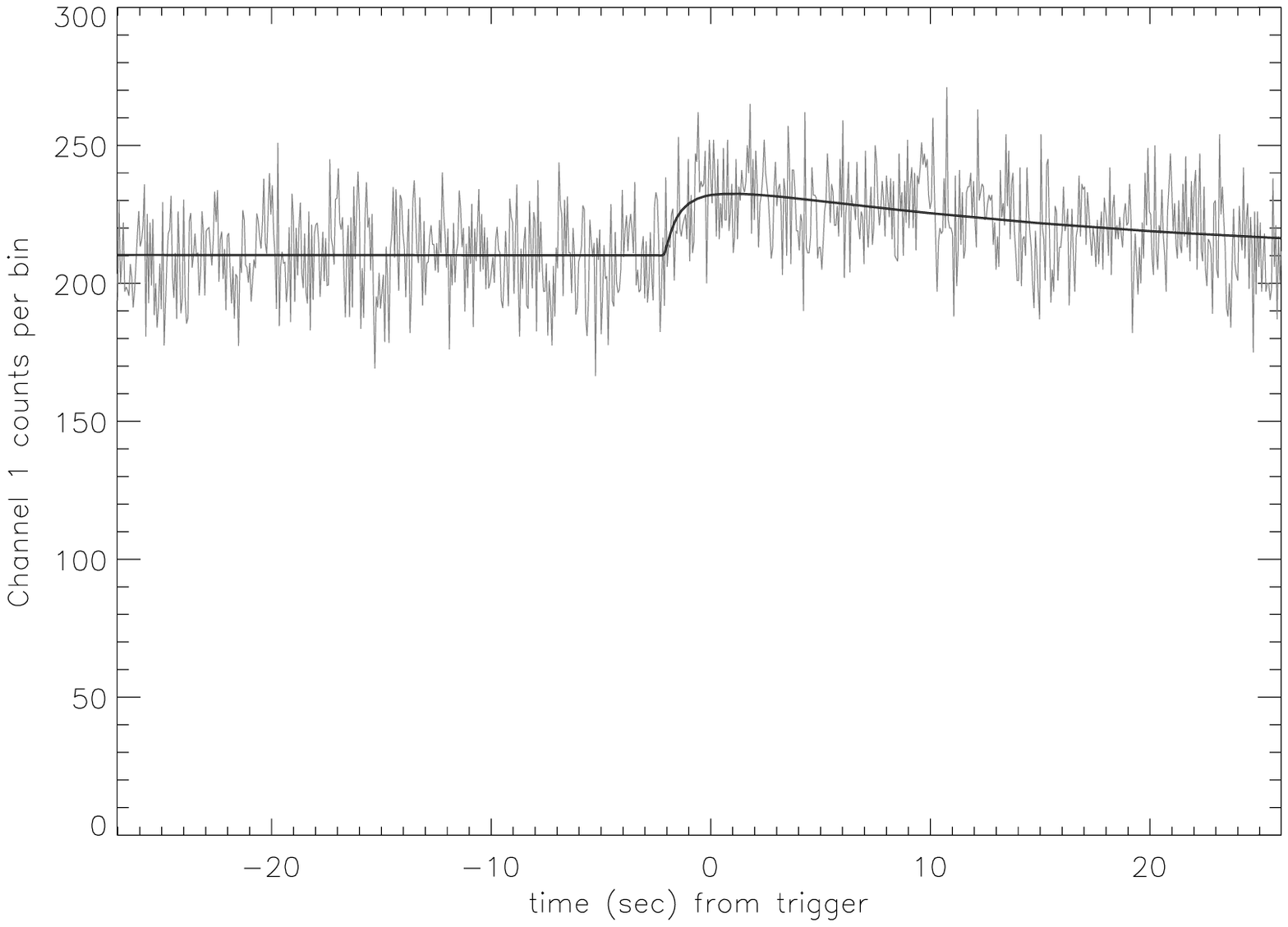}
\caption{Discordant pulse start times in channel 3 (left panel) and channel 1 (right panel) for pulse 1 of BATSE trigger 473. Table \ref{tbl-3} explains why the pulse is subsequently removed from the sample.\label{fig6}}
\end{figure}

\begin{figure}
\plottwo{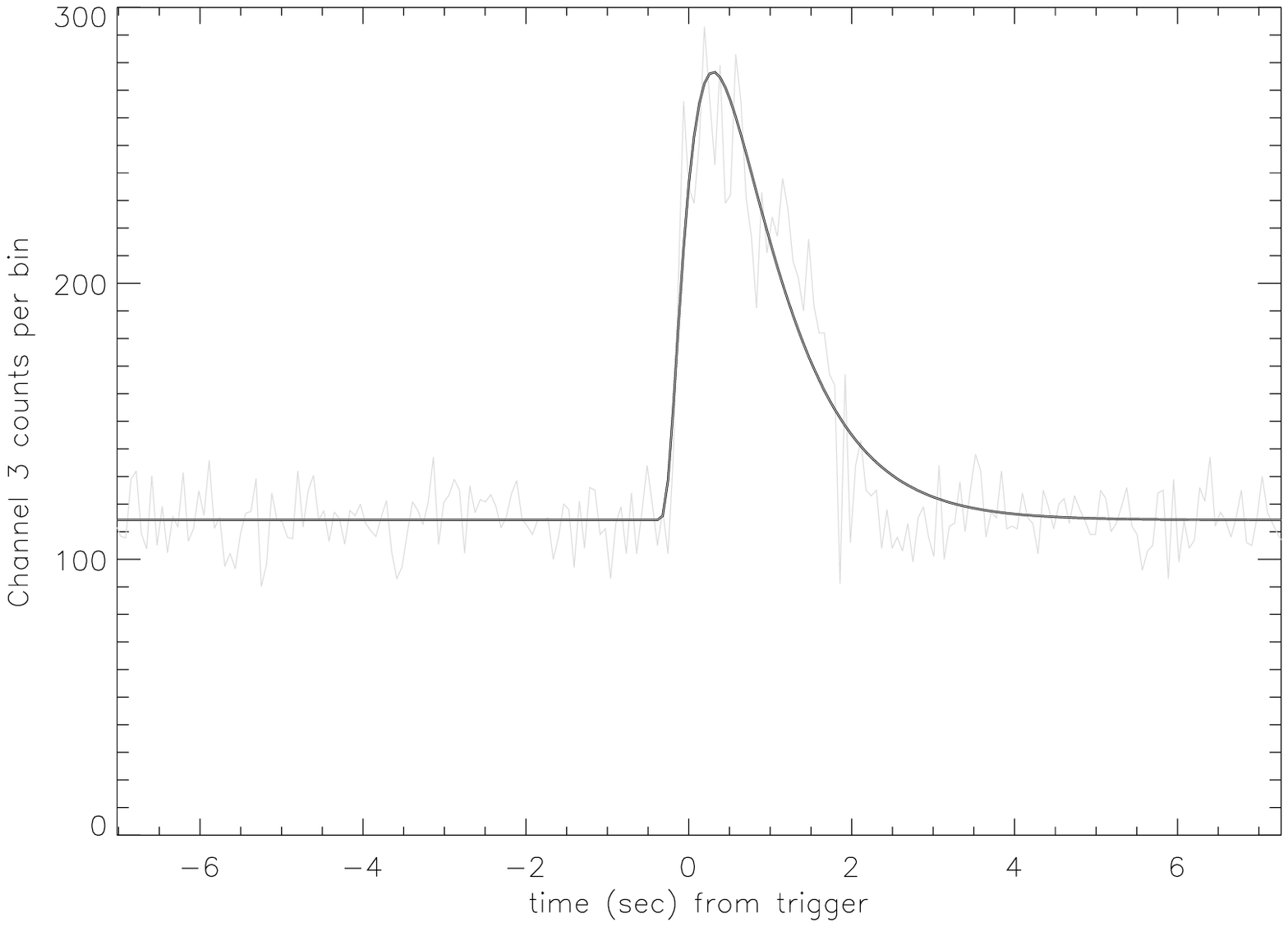}{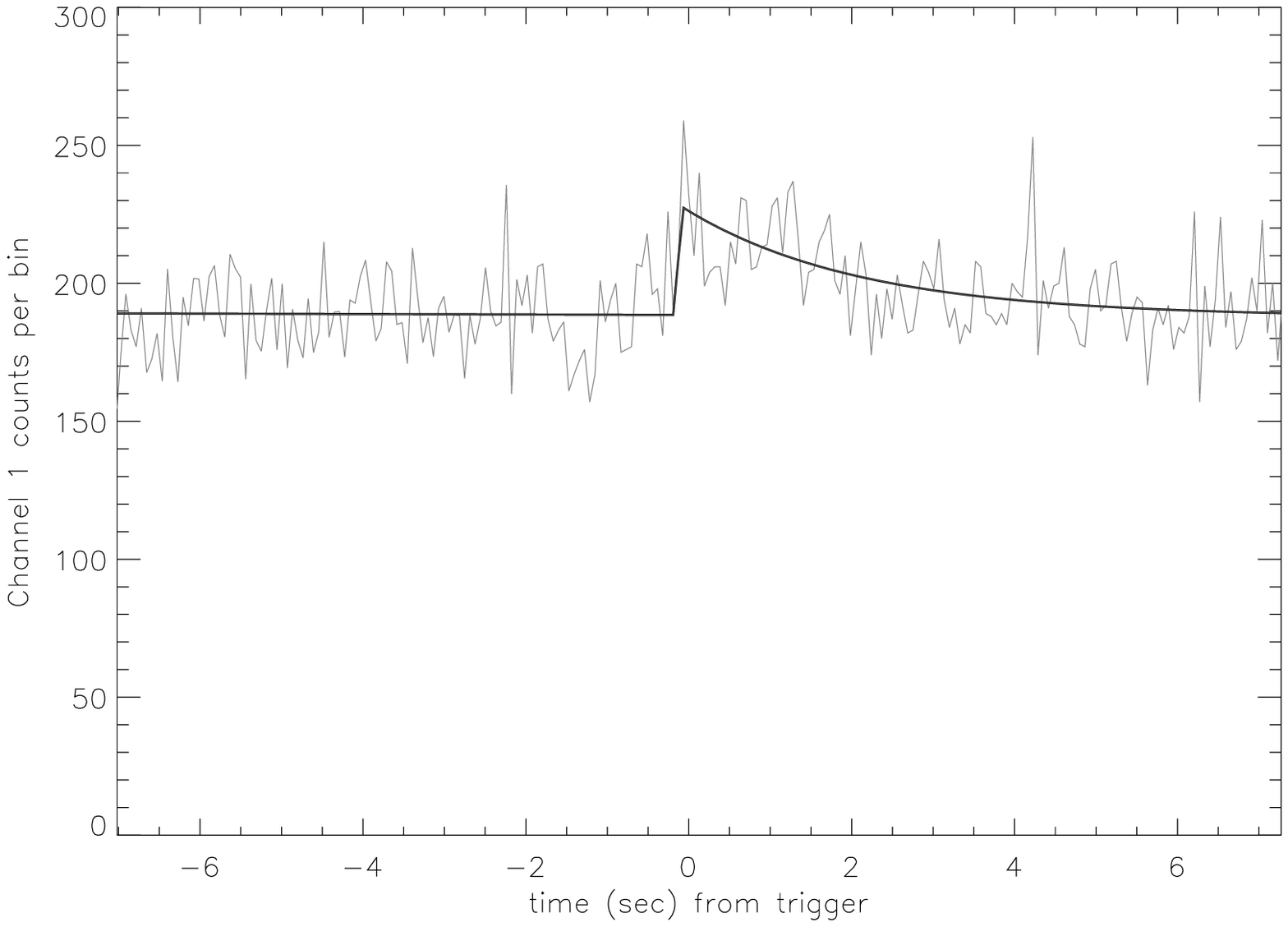}
\caption{Discordant pulse start times in channel 3 (left panel) and channel 1 (right panel) for pulse 1 of BATSE trigger 537. Table \ref{tbl-3} explains why the pulse is subsequently removed from the sample.\label{fig7}}
\end{figure}

\begin{figure}
\plottwo{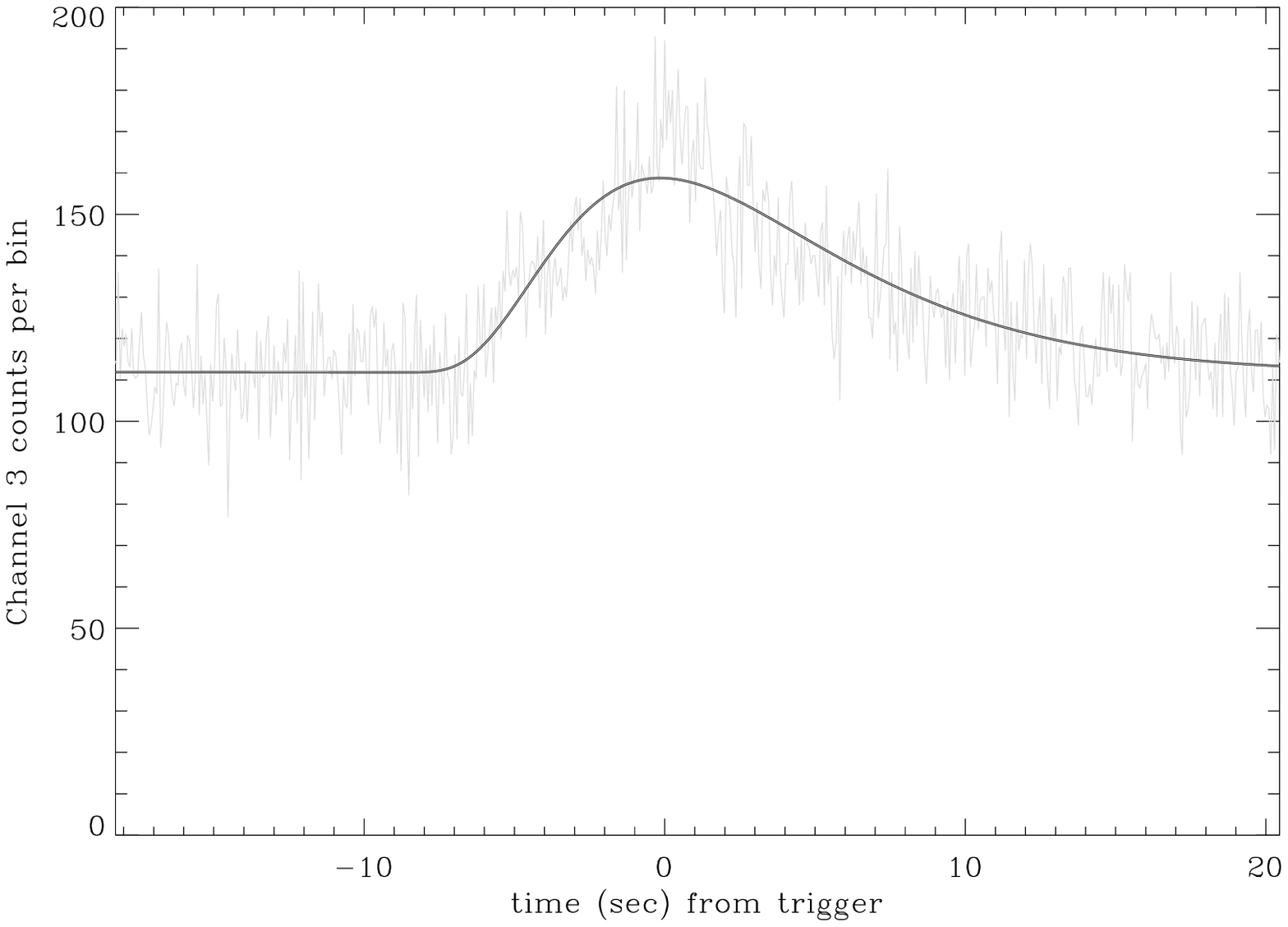}{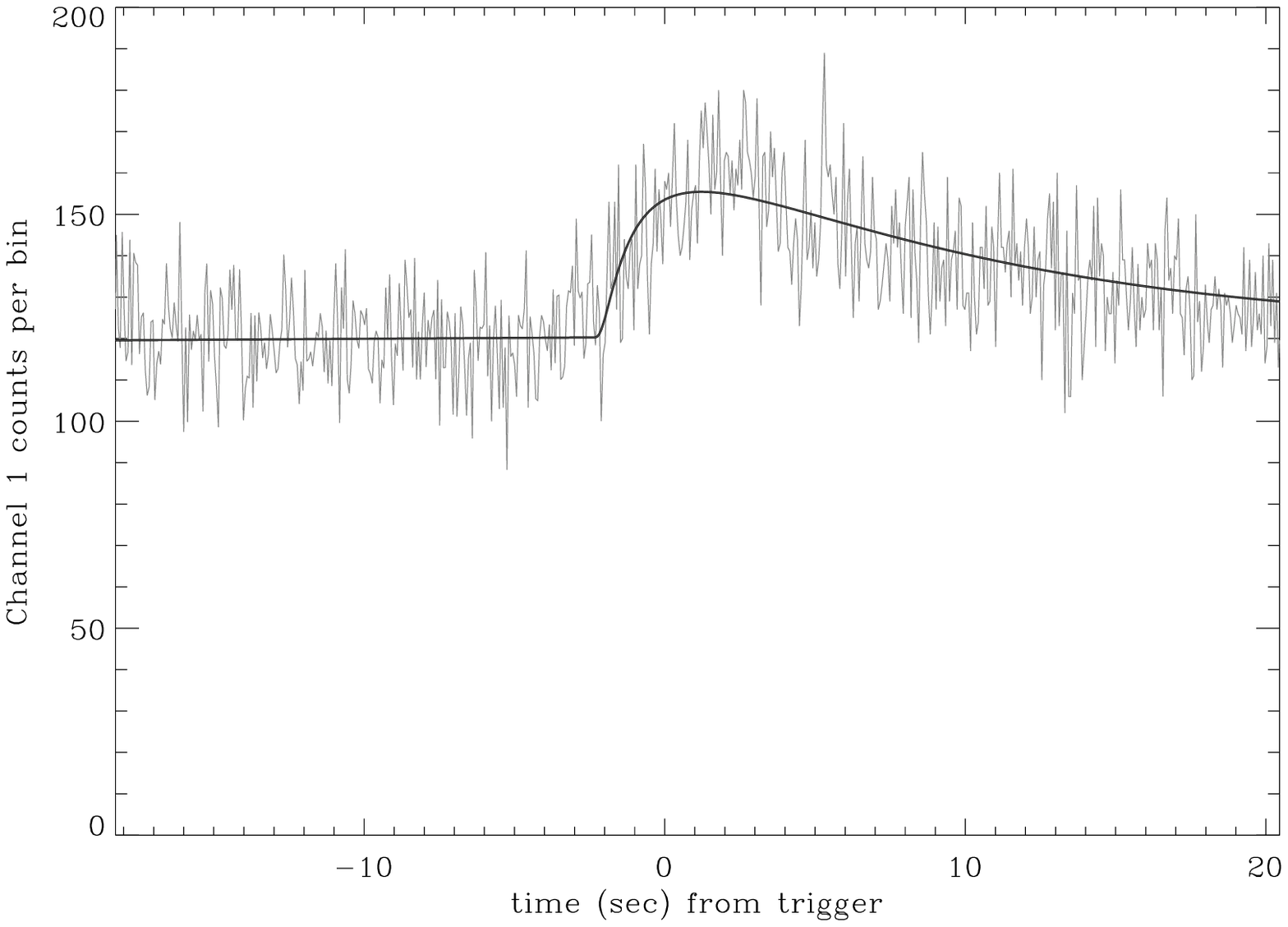}
\caption{Discordant pulse start times in channel 3 (left panel) and channel 1 (right panel) for pulse 1 of BATSE trigger 764. Table \ref{tbl-3} explains why the pulse is subsequently removed from the sample.\label{fig8}}
\end{figure}

\begin{figure}
\plottwo{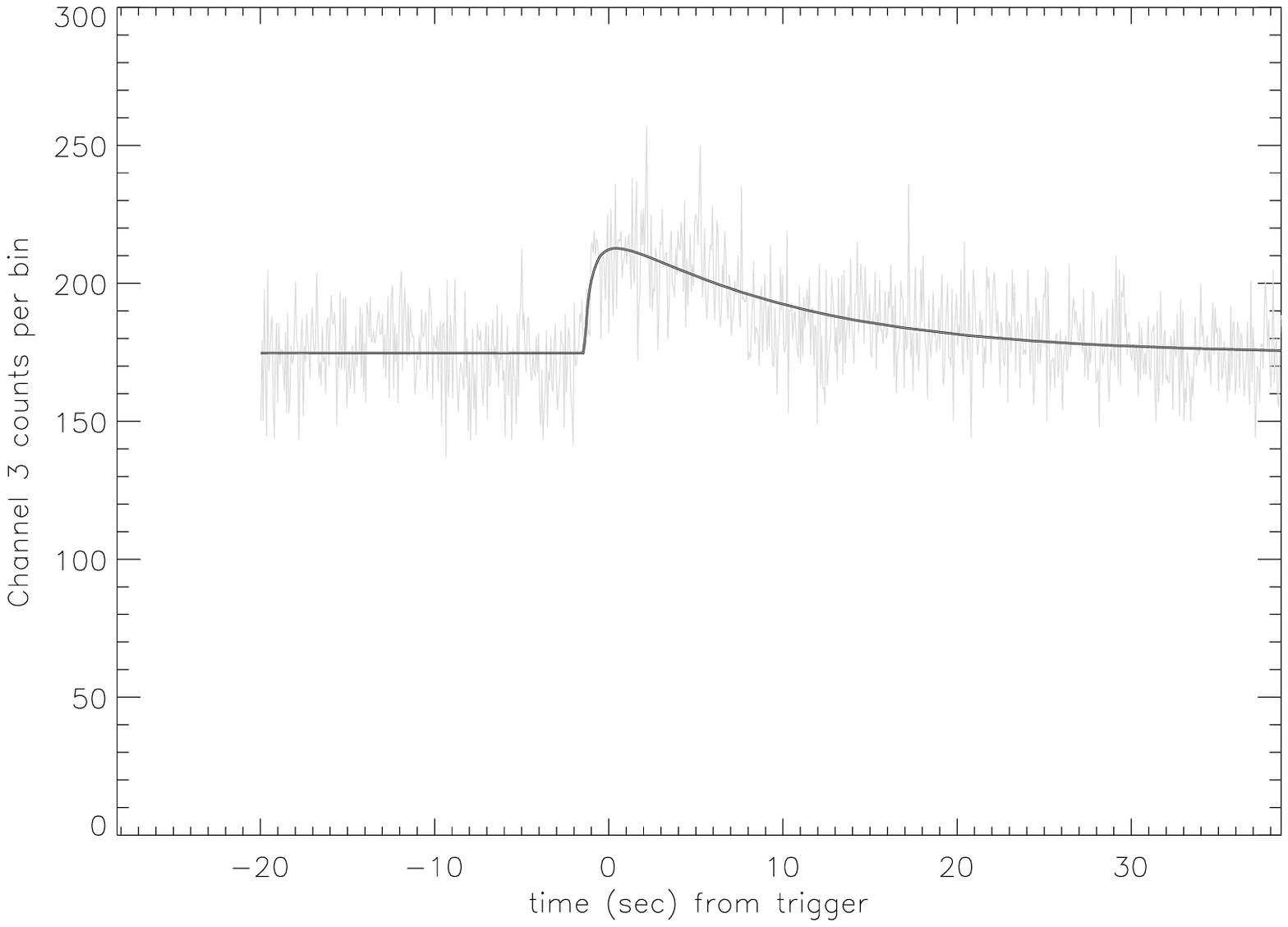}{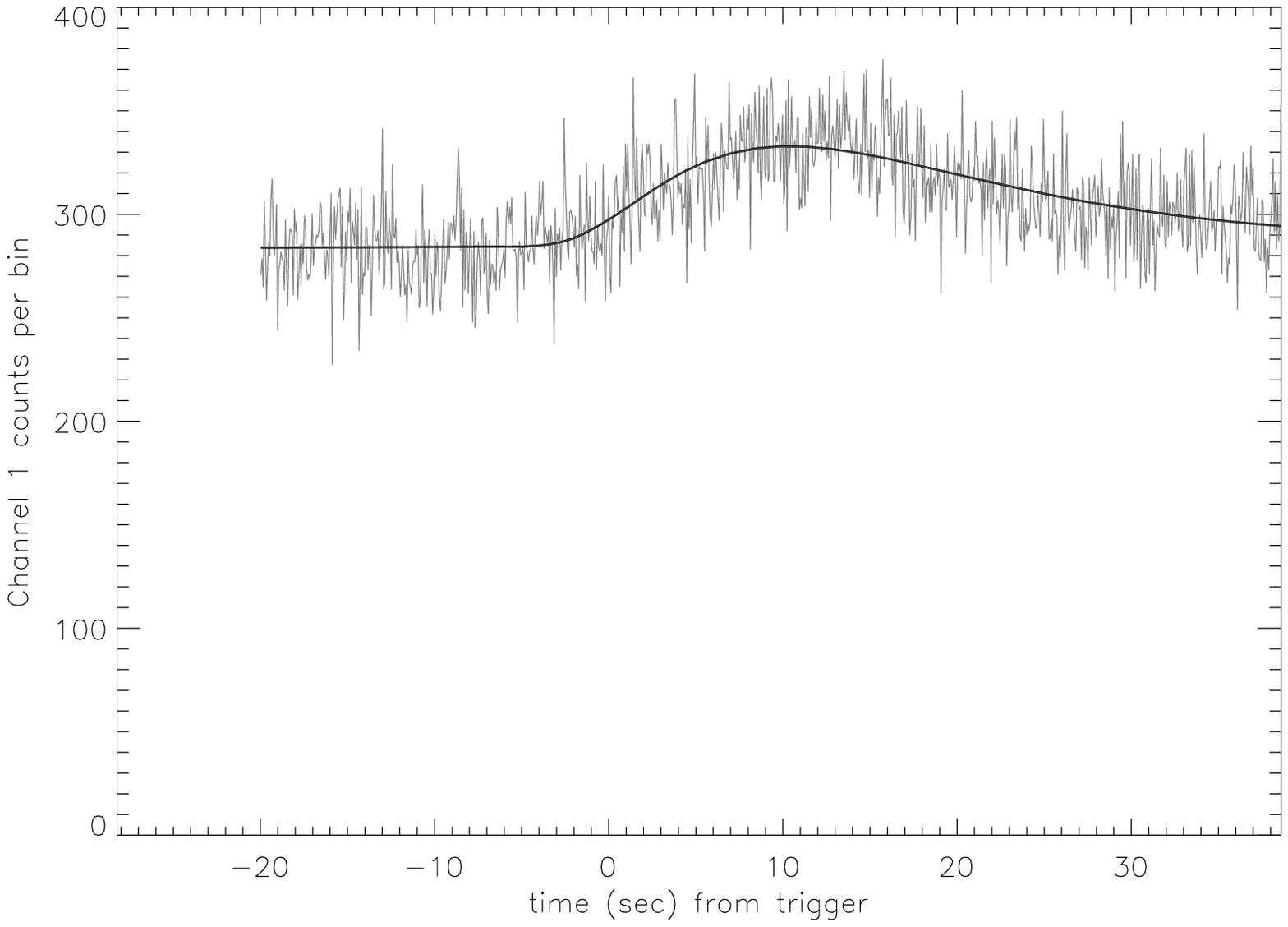}
\caption{Discordant pulse start times in channel 3 (left panel) and channel 1 (right panel) for pulse 1 of BATSE trigger 824. Table \ref{tbl-3} explains why the pulse is subsequently removed from the sample.\label{fig9}}
\end{figure}

\begin{figure}
\plottwo{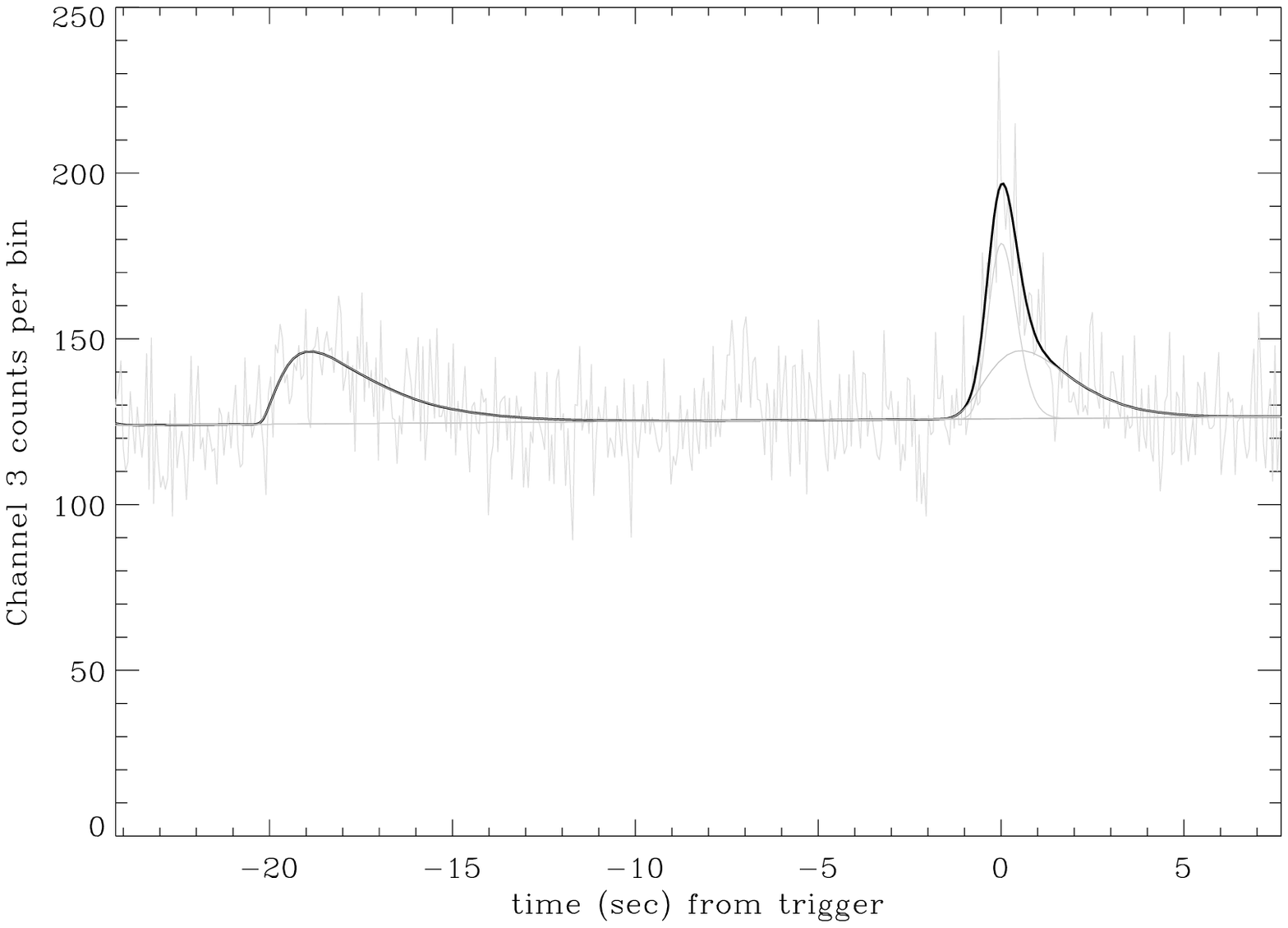}{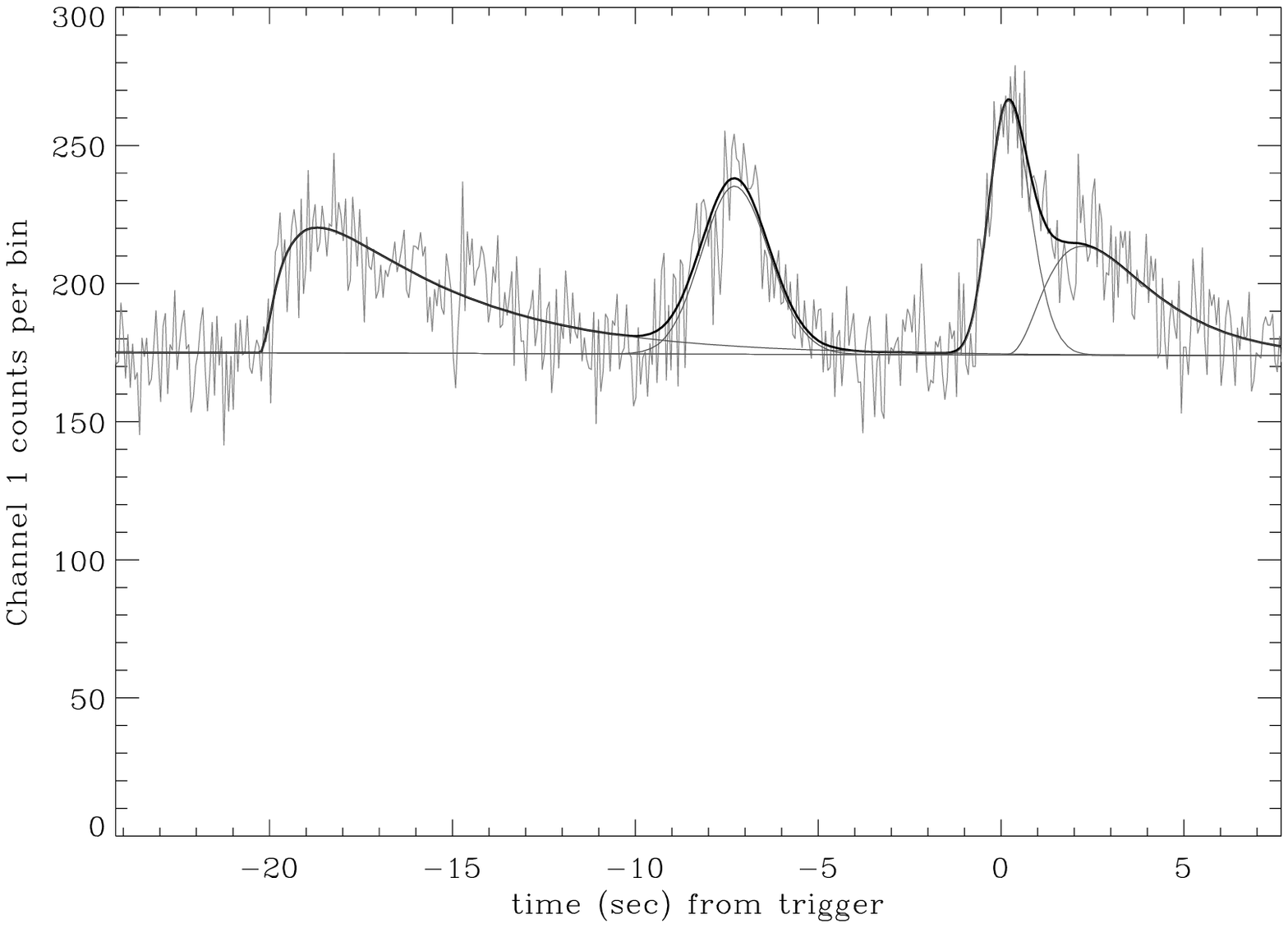}
\caption{Discordant pulse start times in channel 3 (left panel) and channel 1 (right panel) for pulse 2 of BATSE trigger 1042. Pulse 2 is the low amplitude pulse peaking at $t_s=0.55$ s in channel 3 and at $t_s=2.23$ s in channel 1. Table \ref{tbl-3} explains why the pulse is subsequently removed from the sample.\label{fig10}}
\end{figure}

\begin{figure}
\plottwo{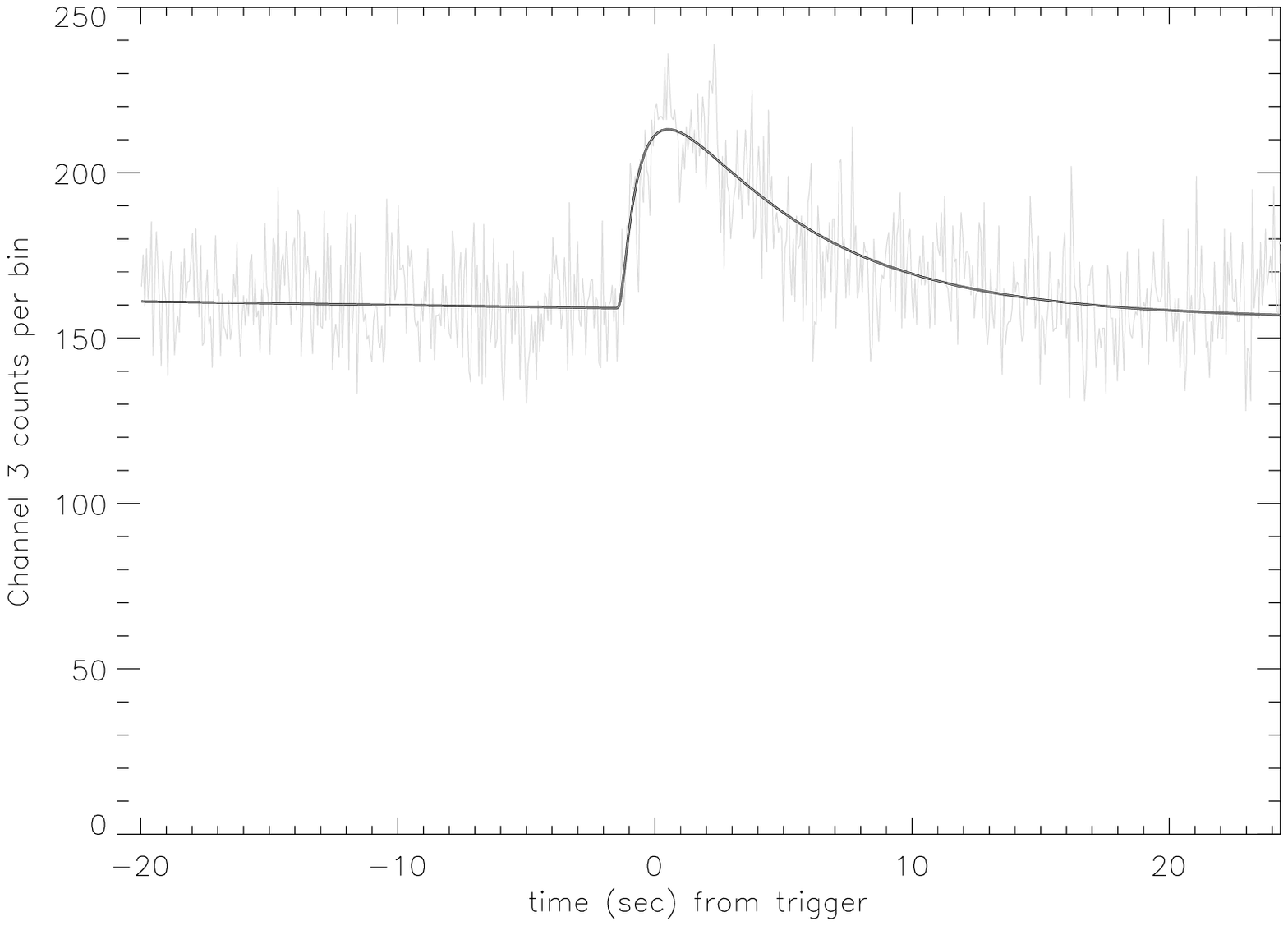}{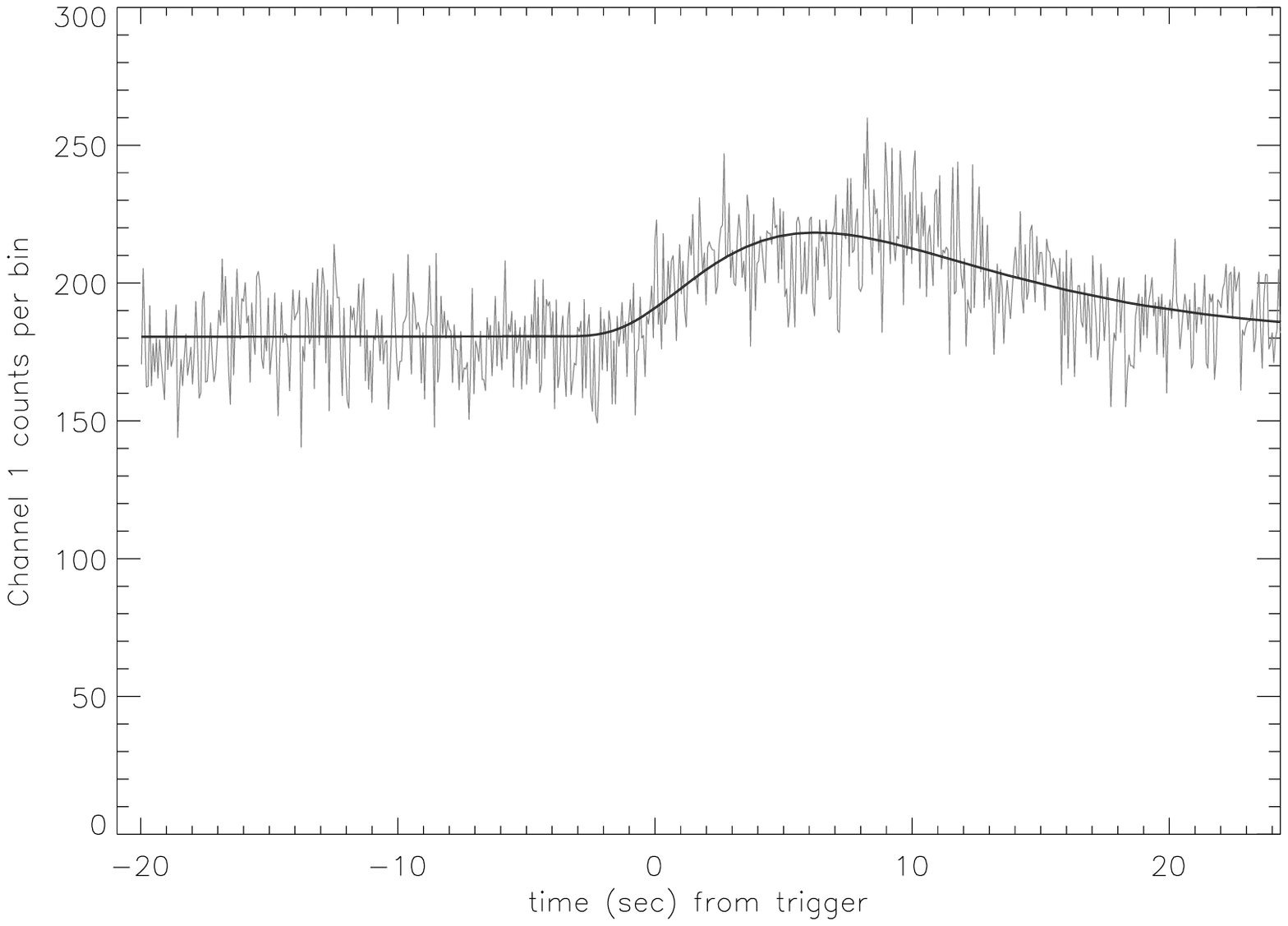}
\caption{Discordant pulse start times in channel 3 (left panel) and channel 1 (right panel) for pulse 1 of BATSE trigger 1561. Table \ref{tbl-3} explains why the pulse is subsequently removed from the sample.\label{fig11}}
\end{figure}

The pulses having $0.032$ s $ \le \sigma_e \le 6$ s and $\chi_i^2 > 5$ are discussed in Table \ref{tbl-3}; these pulses comprise $29$ of the remaining 104 pulses after sample truncation based on $\sigma_e$. Many of these pulses suffer from significant overlap with other pulses in the same burst. Pulse overlap can account for a significant amount of noise in the pulse-fitting process, and can lead to confusion about the pulse start time in different energy channels (particularly if the overlapping pulses have different hardnesses). Eleven of the pulses with large $\chi_i^2$ values have significant pulse overlap that seems to affect start times in different energy channels; these have been excluded. Some fitted pulses exhibit large, persistent variations that might be due to contamination or that might represent faint, unresolved pulses. We have removed roughly half a dozen contaminated pulses. Several pulses are so faint that they exhibit background variations that might be due to noise; these have been removed from consideration as well. A few pulses have unrealistic $\tau_1$ measurements that have affected the $t_s$ uncertainty in the corresponding energy channel; these pulses have also been excluded. Finally, we note the special case of BATSE Trigger 467: this burst appears to have characteristics of a Short GRB with extended emission \citep{nor06}. Our code is not able to uniquely fit the extended emission on this type of burst using a standard pulse model, suggesting that this emission is different from the faint pulses in Long GRBs. We have thus chosen to exclude both the Short GRB pulse and the overlapping extended emission in Trigger 467 from our analysis. Figures \ref{fig2} through \ref{fig11} demonstrate a number peculiar, non-overlapping pulses that we have removed from our analysis. 

\begin{figure}
\plottwo{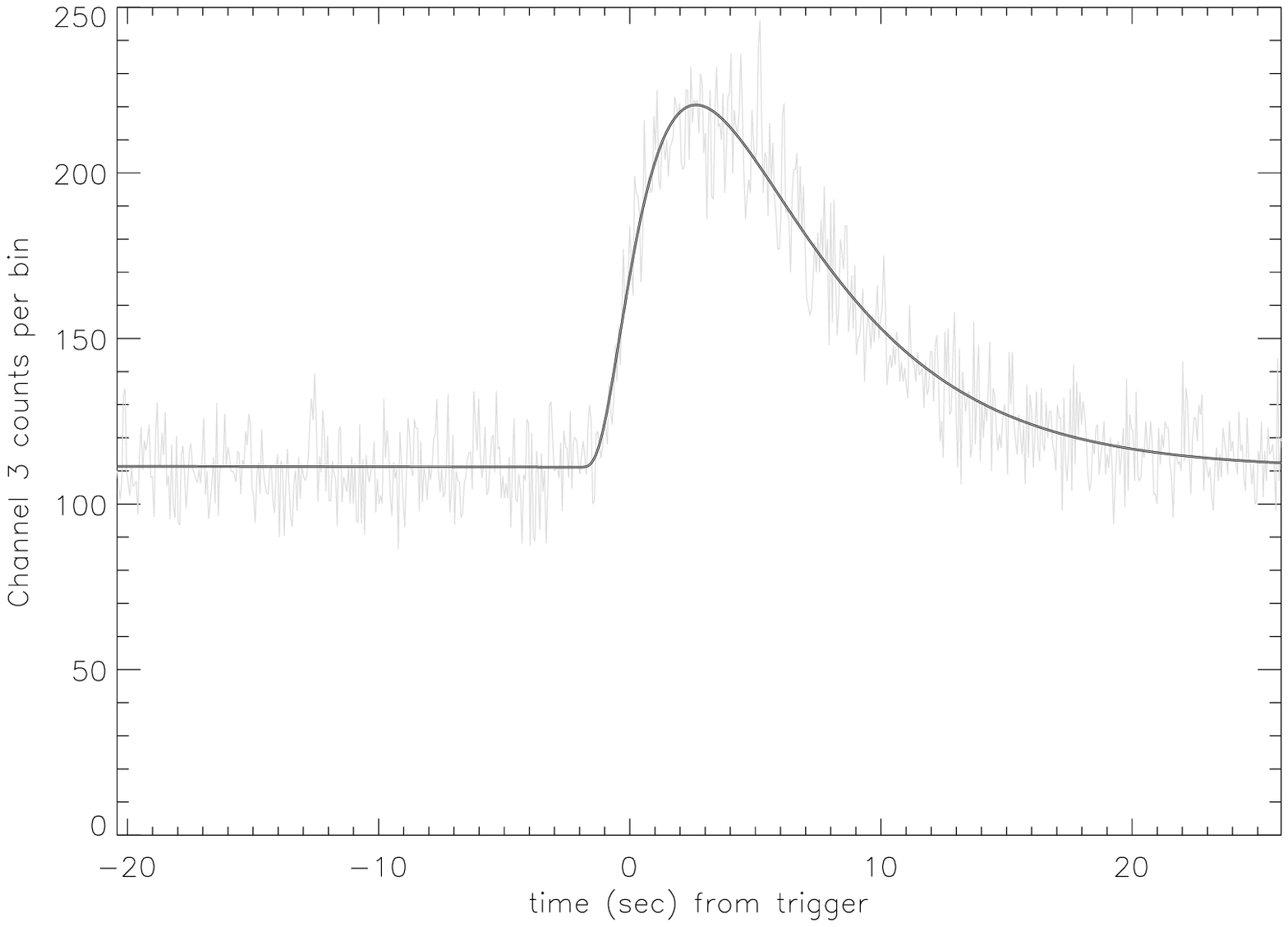}{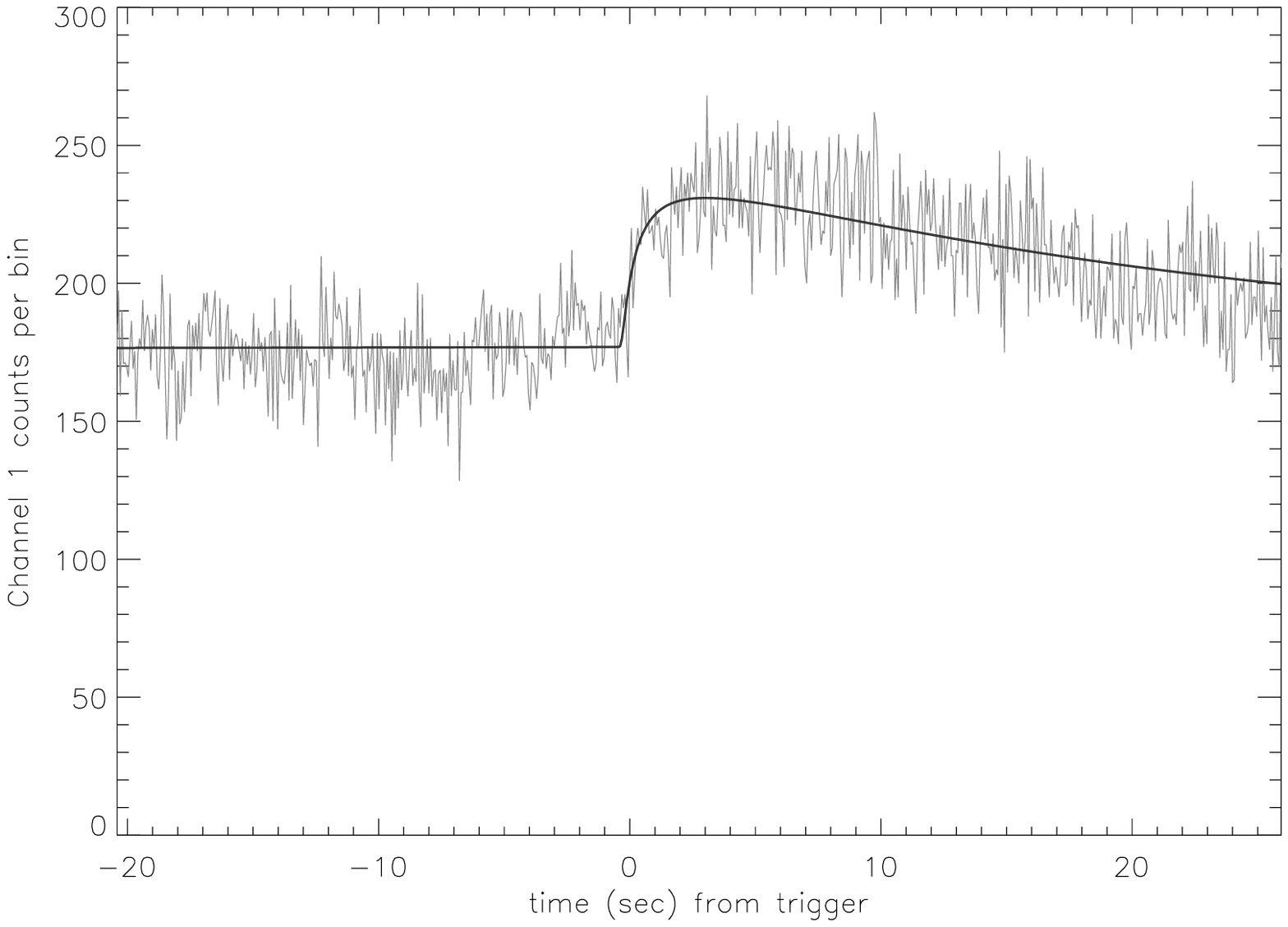}
\caption{Discordant pulse start times in channel 3 (left panel) and channel 1 (right panel) of BATSE trigger 332.  This is one of three Long GRB pulses for which different start times cannot be discounted on the basis of systematic effects.\label{fig12}}
\end{figure}

\begin{figure}
\plottwo{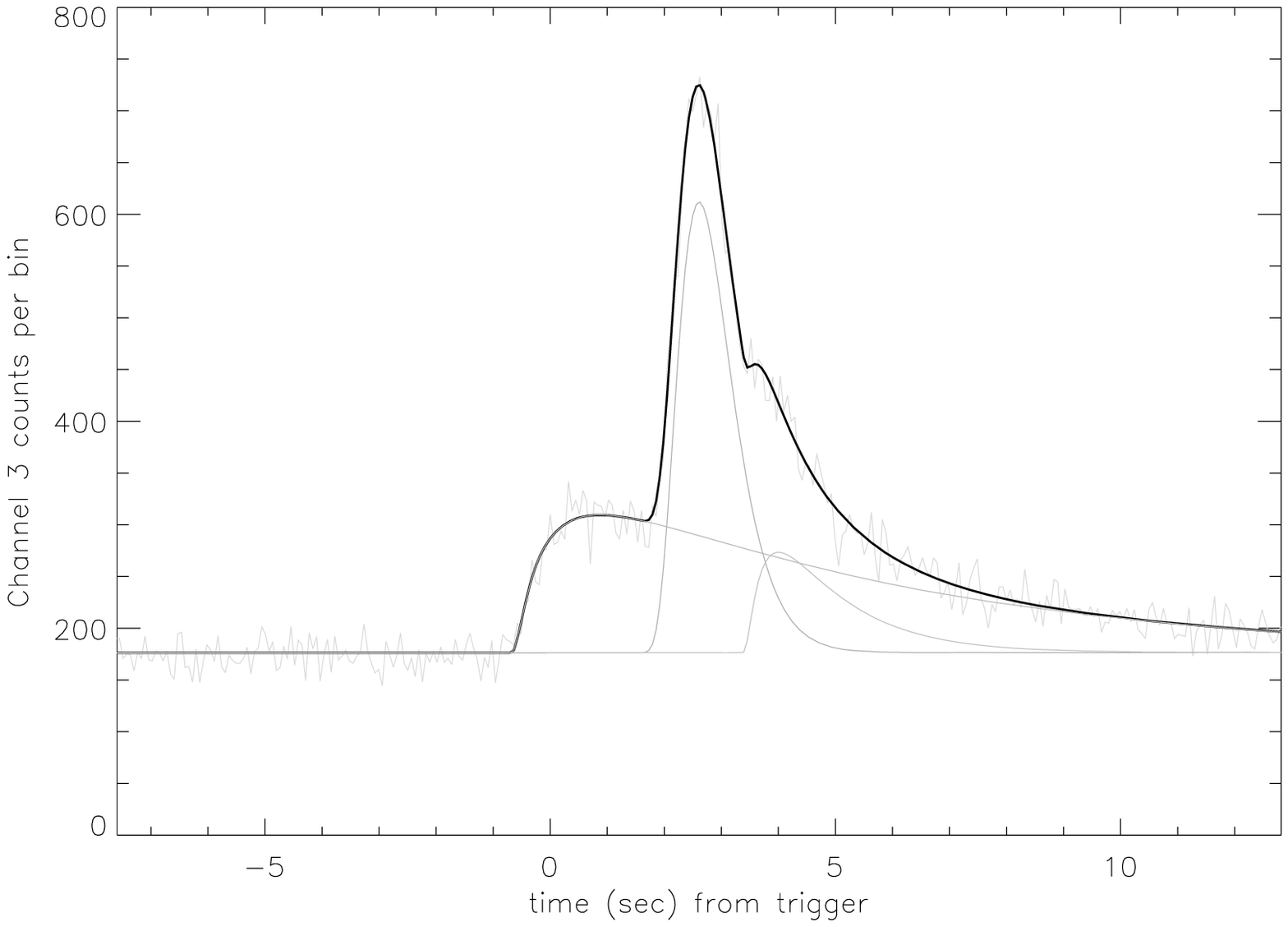}{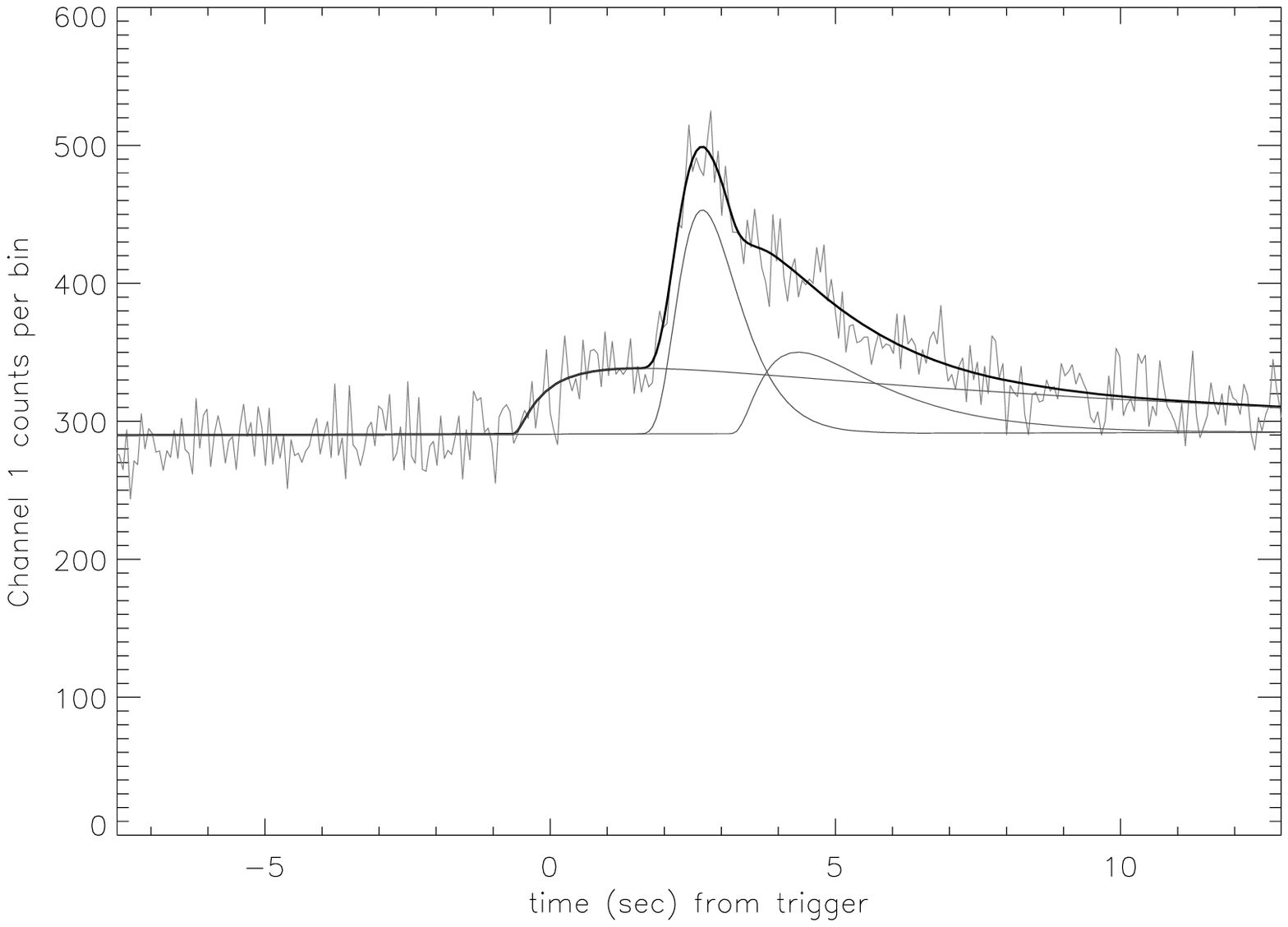}
\caption{Discordant pulse start times in channel 3 (left panel) and channel 1 (right panel) for pulse 1 of BATSE trigger 469. Pulse 1 is the highest-amplitude central pulse, having the shortest duration. This is one of three Long GRB pulses for which different start times cannot be discounted on the basis of systematic effects.\label{fig13}}
\end{figure}

\begin{figure}
\plottwo{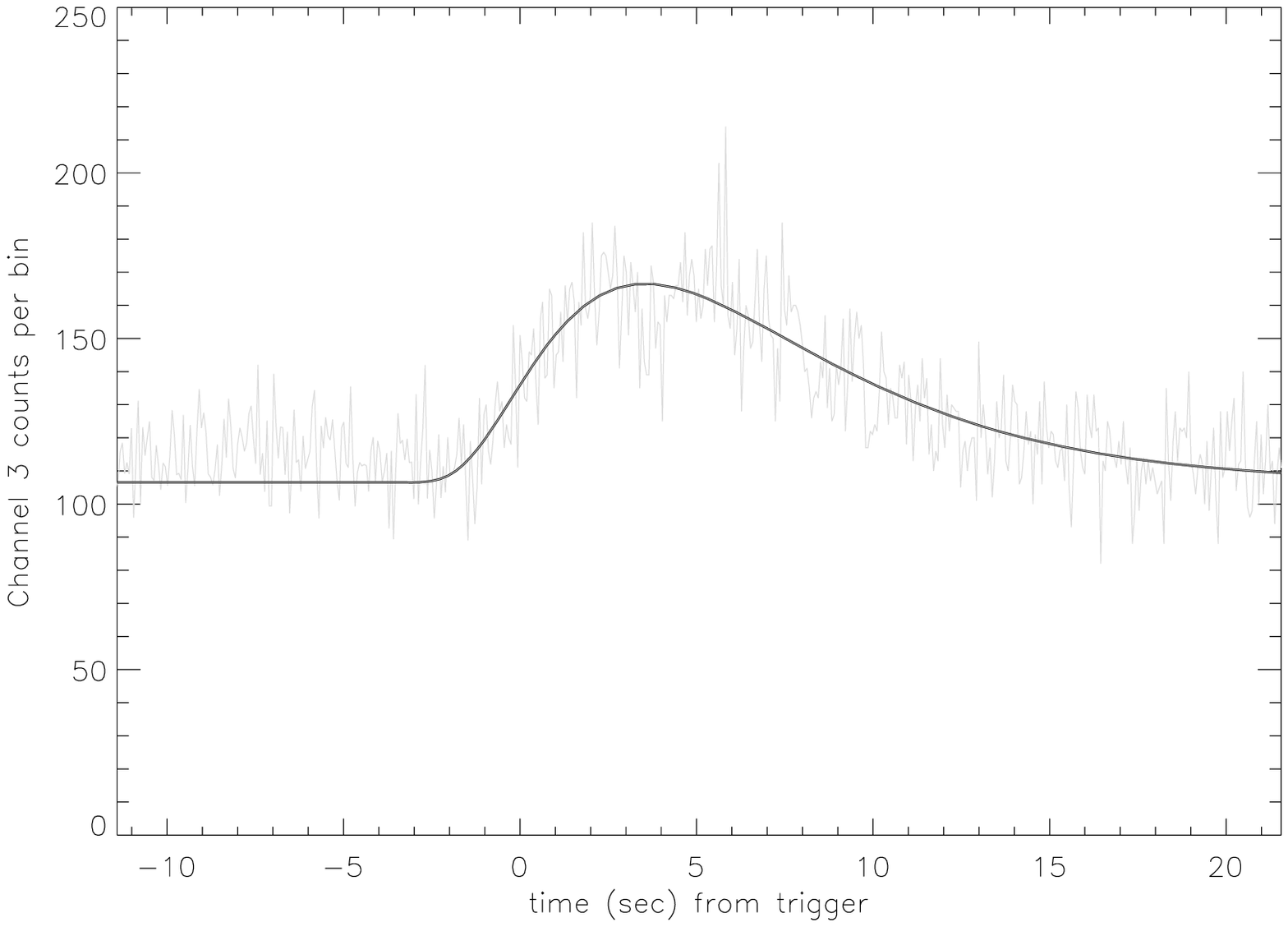}{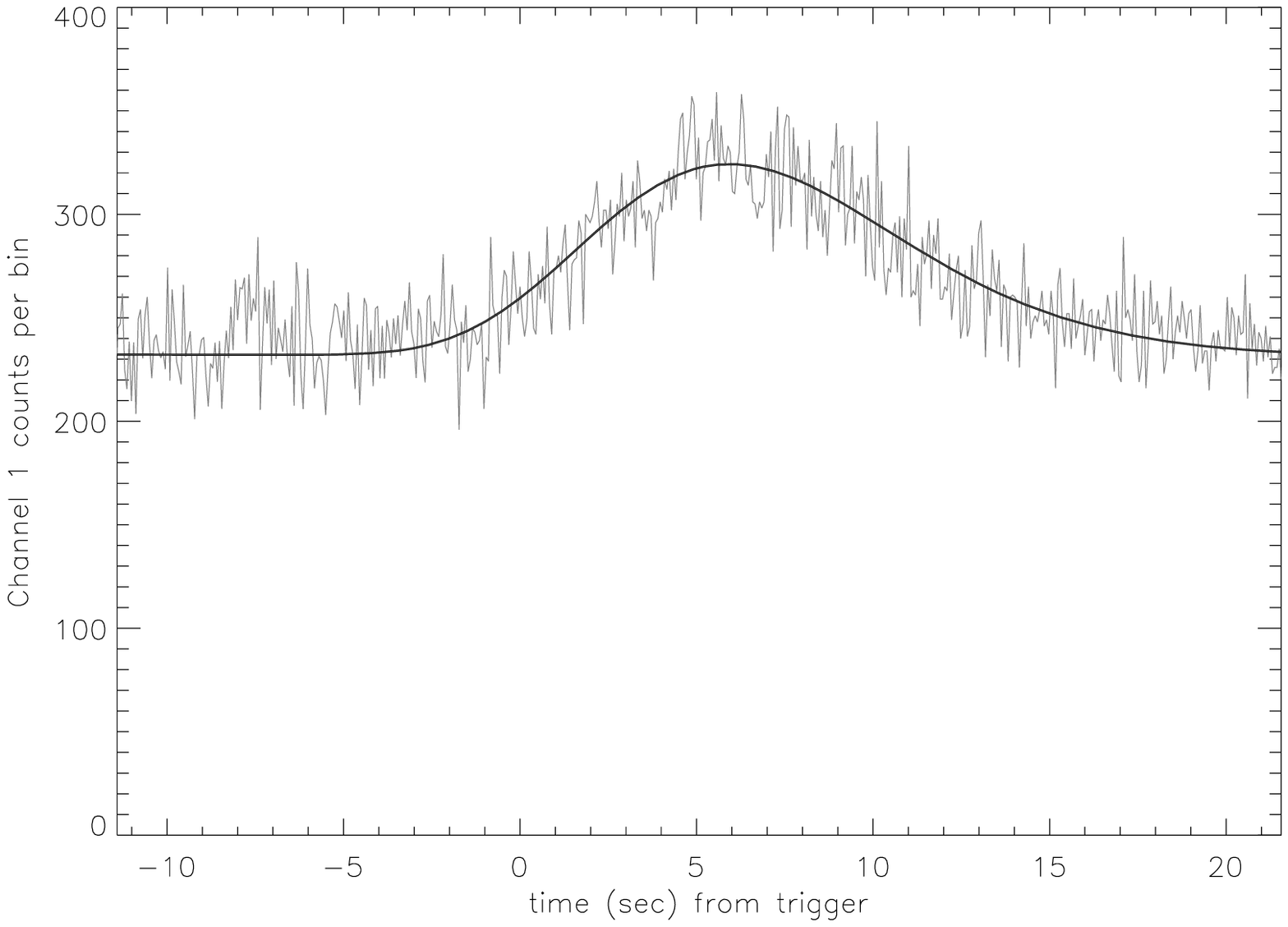}
\caption{Discordant pulse start times in channel 3 (left panel) and channel 1 (right panel) for pulse 1 of BATSE trigger 1200. This is one of three Long GRB pulses for which different start times cannot be discounted on the basis of systematic effects.\label{fig14}}
\end{figure}

We cannot easily exclude 3 of the 29 pulses shown in Table \ref{tbl-3}. The first pulses of Triggers 332, 469, and 1200 all have high signal-to-noise and/or no pulse overlap, and are considered to have valid start time measurements. These pulses, shown respectively in Figures \ref{fig12}, \ref{fig13}, and \ref{fig14}, have been included in our subsequent analysis.

The resulting $\chi_i^2$ distribution, after removal of pulses having start times possibly marred by the systematic effects described above, is plotted in Figure \ref{fig14}. This distribution is consistent with a theoretical one;  the probability of obtaining a value greater than or equal to a reduced of $\chi^2 = 1.6$ is $p=0.15$. The Pulse Start Conjecture thus appears to be valid for this sample (including discrepant pulses in Triggers 332, 469, and 1200) because the measured variations in pulse start times are consistent with the formal uncertainties. The three pulses in triggers 332, 469, and 1200 may violate the Pulse Start Conjecture. They may, however, also represent the tail of the pulse start time $\chi_i^2$ distribution; this tail might be extended due to the non-Poisson nature of the gamma-ray background. 

\begin{figure}
\plotone{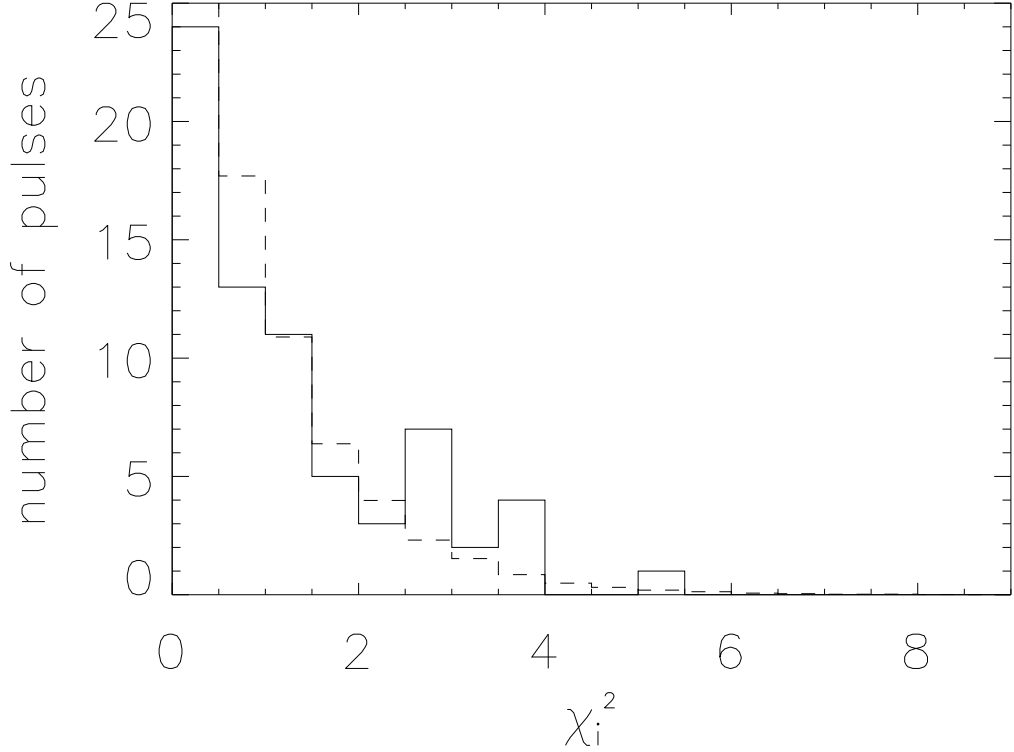}
\caption{The distribution of $\chi_i^2$ (solid histogram) for combined Long GRB pulse start times across energy channels 1, 2, and 3 (equation 2) is compared to the expected distribution (dotted histogram) after removal of pulses having discordant start time uncertainties. The probability of obtaining a value greater than or equal to a reduced of $\chi^2 = 1.6$ is $15\%$, suggesting that pulses with discordant start time uncertainties have largely been removed. The pulses removed from consideration have unrealistically small ($\sigma_e < 0.032$ s) or unrealistically large ($\sigma_e > 6$ s) start time uncertainties because these uncertainties are related to poorly-constrained $\tau_1$values. Also removed are pulses having large $\chi_i^2$ values that can be explained by systematic errors (see Table \ref{tbl-3} and Figures \ref{fig3} through \ref{fig13}. Not shown on the plot (but included in the analysis) are discrepant GRB pulses BATSE 332 p01, BATSE 469 p01 and BATSE 1200 p01. \label{fig15}}
\end{figure}

The start time variations for the retained pulses are
\begin{equation}
\sigma_i = \sqrt{\frac{1}{n-1}\sum_{e=1}^n (t_{s ; e}-\bar{t}_s)^2},
\end{equation}
and these can be used to place limits on the Pulse Start Conjecture. The distribution of $\sigma_i$ is shown in Figure \ref{fig15}. Although a few of the pulses have relatively unconstrained start times ($\sigma_i \ge 1$) s, most have start times that are consistent with the Pulse Start Conjecture to within $0.4$ s, and $\approx 40\%$ of them are consistent with the Conjecture to within $0.2$ s (although our pulse selection criteria do not allow us to place constraints any tighter than $0.032$ s). Thus, we believe that the Pulse Start Conjecture is typically valid for Long GRB pulses within a 0.4 s window, with tighter constraints indicated in a few cases.

\subsubsection{Short GRB Pulses}

It is much more difficult to constrain the start times of Short GRB pulses than Long ones. Short GRB pulses are typically much shorter and more intense than Long GRB pulses \citep{hak09}, and their pulses typically occupy only a small number of 64 ms bins. Furthermore, Short GRB pulses are typically hard, often having limited flux in channel 1 and additional flux in relatively insensitive channel 4. As a result, many Short GRB pulses have unacceptably small or large start time uncertainties that are not usable by the start time analysis. 

We have repeated the start time analysis to test the Pulse Start Conjecture for the sample of 49 Short GRB pulses, again limiting our data to BATSE channels 1, 2, and 3, and truncating the data as described in the previous section ($0.032$ s $ \le \sigma_e \le 6$ s and $\chi_i^2 > 5$) because these constraints were applicable to Long GRB pulses. Unfortunately, only six pulses survive our pulse rejection criteria. One of these pulses (pulse 3 of Trigger 603) has a large $\chi_i^2$, but also has low signal-to-noise and overlaps another pulse. Excluding it leaves us with a sample of 5 Short GRB pulses. This is, unfortunately, too small to reliably test the Pulse Start Hypothesis, although the data are suggestive that the hypothesis is valid within a very short window ($\approx 0.1$ s). Pulse fits are needed for data with better time resolution than 64 ms to test the Pulse Start Hypothesis for Short GRB pulses.

\section{Discussion}

Our analysis generally supports the Pulse Start hypothesis, namely that GRB pulses begin simultaneously at all energies, within our formal uncertainty of accurately identifying these start times. The model-dependent approach we have used to study this hypothesis allows us to place limits on the time interval during which pulse emission begins; typically this is $\sigma_{t_s} < 0.4$ s for pulses in Long GRBs ($\sigma_{t_s} < 0.2$ s for $40\%$ of the pulses) and perhaps $\sigma_{t_s} < 0.1$ s for pulses in Short GRBs.

These may sound like fairly broad, unconstrained time intervals within which the pulse begins. However, GRB prompt emission suffers from small number counting statistics; gamma-rays are not plentiful in GRB prompt emission, and the gamma-ray background is noisy and often varies in a non-Poisson way. We have reduced the size of our sample by collecting photons in discrete energy channels, grouping them into time bins, then fitting these bins with the pulse model. The pulse start time is then the parameterized time at which emission increases from being undetectable to the first moment at which it is detectable. Considering these limitations, we believe that the constraints found in this study are realistic.

The results thus strongly suggest that pulse emission does not preferentially build up at either low energies or high energies at the beginning of GRB pulses; the emission appears to result from a simultaneous and rather rapid energy deposition. When considered in conjunction with pulse spectral evolution observations, this has repercussions that can be applied usefully to theoretical modeling. GRB pulses are known to exhibit hard to soft spectral evolution. In terms of the Band spectral parameters \citep{ban93}, this evolution represents a decay of the peak energy $E_p$ (e.g. \cite{kan06, pen09}. In terms of pulse modeling, pulses peak first at high energies, then subsequently at lower energies. Thus, pulse evolution appears to indicate a softening of the photon distribution (and perhaps cooling) from the moment the pulse is observed until it ends. Remarkably, it also indicates that the pulse peak intensity is not a critically important pulse attribute since this attribute depends on the energy response of the instrument, the amount of energy radiated by the pulse, and the GRB redshift; the pulse peak flux represents only the flux across the instrumental bandpass at the time that a pulse peaks at one particular energy.

The observation that a GRB is composed of many pulses, and that pulse duration, lag, luminosity, and hardness are all related suggests that the amount of energy initially deposited drives the pulse evolution. GRBs with many pulses have several energy injections, and the amount of energy injected strongly influences all of the pulse observables. Since the order of pulses in a GRB does not always proceed from hard to soft, this implies that the energy injection is made up of independent events.

The results demonstrated here suggest that pulse modeling can be greatly simplified, because several of the free parameters in the \cite{nor05} pulse model are not really ``free;'' there appears to be only one pulse start time, as opposed to one per energy channel fit. To further support this statement, we compare the pulse start time `lag' (the difference between the channel 3 and channel 1 start times) with the durations of Long burst pulses having well-measured start times (e.g., those in Figure \ref{fig15}). The results, plotted in Figure \ref{fig17}, indicate that these two parameters are uncorrelated (the Spearman Rank Order significance is $0.266$ that a random distribution has a smaller coefficient than the measured value of $-0.132$). This result is contrary to the high level of correlation that exists between duration and other energy-dependent pulse parameters (e.g.\ lag and spectral hardness) found by \cite{hak09}, and supports the idea that pulse start times are simultaneous rather than energy-dependent.

\section{Acknowledgments}

We thank our referee, Jerry Bonnell, for his helpful and insightful comments and suggestions. We are also grateful to Rob Preece, Jay Salmonson, Jay Norris, and Tom Loredo for helpful discussions on aspects of pulse modeling.  The material presented here is based upon work supported by NASA under award No. GRNASNNX06AB43G and through the South Carolina NASA Space Grant program.

\clearpage

\clearpage

\clearpage

\clearpage

\clearpage

\clearpage

\clearpage

\clearpage

\clearpage

\clearpage

\begin{figure}
\plotone{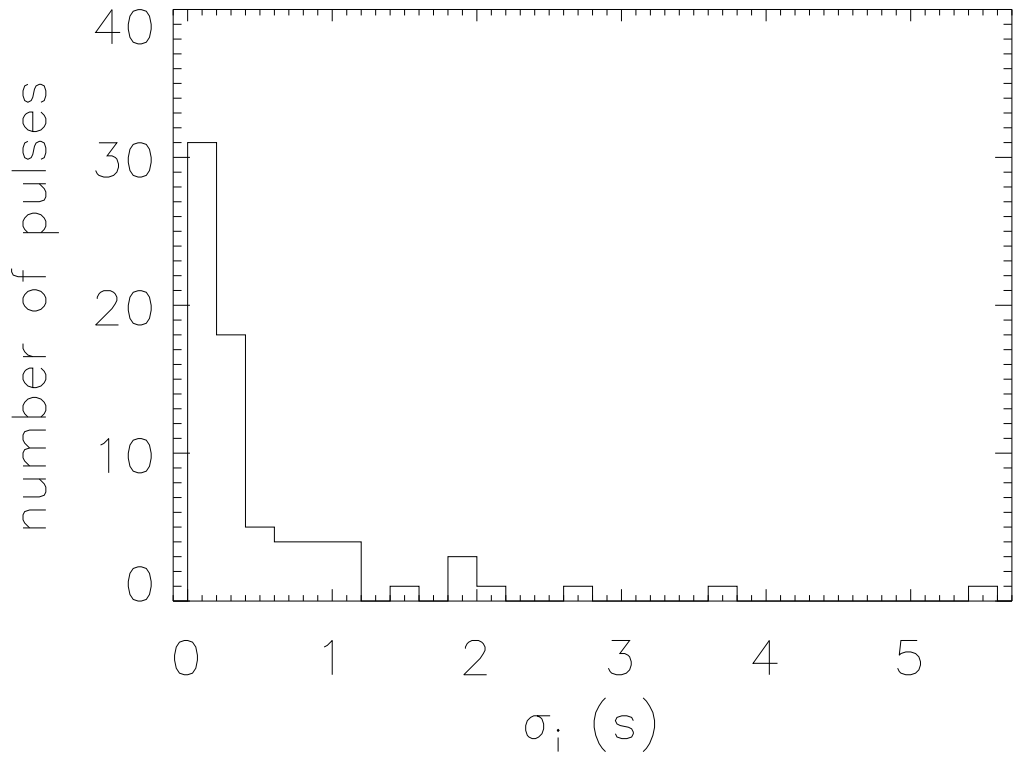}
\caption{The distribution of $\sigma_i$ for the $i$ Long GRB pulses shown in Fig. \ref{fig15}; these pulses appear to satisfy the Pulse Start Conjecture within the plotted $\sigma_i$. \label{fig16}}
\end{figure}

\clearpage

\begin{figure}
\plotone{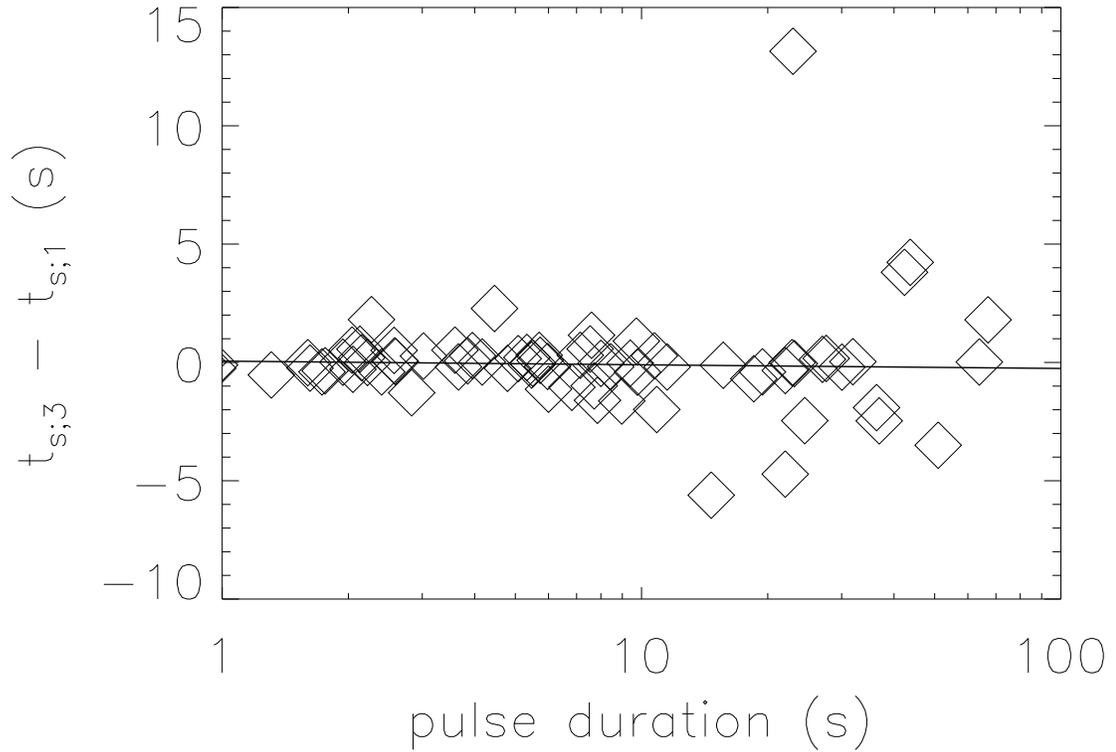}
\caption{Comparison of start time `lags' ($t_{s; 3}-t_{s; 1}$) vs.\ durations for the Long GRB pulses included in Figure \ref{fig15}. There is no correlation between these two parameters such as the ones found between duration and energy-dependent pulse parameters such as lag and spectral hardness: this result supports the idea that pulses start near-simultaneously in all energy channels, rather than in an energy-dependent way. \label{fig17}}
\end{figure}

\clearpage

\begin{deluxetable}{ccccccccccc}
\tabletypesize{\scriptsize}
\rotate
\tablecaption{Start times for some Long GRB pulses. Start times are given in seconds since the trigger, and uncertainties are given in seconds. An asterisk (*) indicates insufficient signal in the energy channel to fit the pulse. \label{tbl-1}}
\tablewidth{0pt}
\tablehead{
\colhead{BATSE Pulse} & \colhead{$t_{s;\rm all}$} & $\sigma_{t_{s;\rm all}}$ & \colhead{$t_{s;1}$} & \colhead{$\sigma_{t_{s;1}}$} &
\colhead{$t_{s;2}$} & \colhead{$\sigma_{t_{s;2}}$} & \colhead{$t_{s;3}$} & \colhead{$\sigma_{t_{s;3}}$} & \colhead{$t_{s;4}$} &
\colhead{$\sigma_{t_{s;4}}$}
}
\startdata
0105 p01 & -1.41 & 0.214 & -15.38 & 16.89 & -28.81 & 48.13 & -11.94 & 10.61 & -0.74 & 1.49\\
0105 p02 & 2.86 & 0.02 & 2.92 & 0.04 & 2.78 & 0.04 & 2.85 & 0.03 & 4.18 & 7.28\\
0105 p03 & -21.23 & 63.96 & -0.36 & 2.14 & -45.85 & 354.6 & -47.15 & 450.48 & -22.95 & 487.34\\
0111 p01 & -17.01 & -0.87 & -16.13 & 1.57 & -17.53 & 1.37 & -19.88 & 2.34 & -10.83 & 1.64\\
0130 p01 & -120.15 & 101.48 & -123.03 & 215.72 & -232.03 & 524.09 & -125.64 & 157.88 & * & *\\
0130 p02 & -17.44 & 108.68 & -13.08 & 216.94 & -42.88 & 943.17 & -18.86 & 122.88 & * & *\\
0130 p03 & 4.07 & 0.46 & 3.79 & 1.47 & 4.20 & 0.56 & 4.00 & 0.62 & * & *\\
0130 p04 & 5.00 & 1.72 & 5.04 & 4.25 & 5.15 & 2.62 & 3.44 & 3.57 & * & *\\
0130 p05 & 9.41 & 9.36 & 9.27 & 20.89 & 9.59 & 14.3 & 9.69 & 13.32 & 9.65 & 16.86\\
0130 p06 & 14.75 & 0.32 & 14.76 & 0.71 & 13.89 & 1.01 & 15.37 & 0.20 & 14.77 & 1.27\\
0130 p07 & 13.1 & 0.45 & 11.50 & 3.70 & 13.22 & 0.49 & 13.31 & 0.52 & * & *\\
0130 p08 & 24.87 & 0.12 & 25.02 & 0.18 & 20.24 & 12.88 & 25.09 & 0.05 & 23.78 & 8.30\\
0130 p09 & 27.37 & 0.59 & 26.35 & 5.45 & 22.89 & 14.00 & 28.42 & 1.08 & * & *\\
0130 p10 & 35.00 & 0.06 & 35.12 & 0.09 & 34.48 & 0.16 & 35.02 & 0.09 & * & *\\
0130 p11 & 26.10 & 0.97 & 20.38 & 3.81 & 23.29 & 2.53 & 29.18 & 0.86 & * & *\\
0130 p12 & 40.05 & 0.22 & 40.30 & 0.30 & 39.86 & 0.51 & 40.05 & 0.16 & 38.69 & 12.10\\
0130 p13 & 34.47 & 26.48 & 34.08 & 72.14 & 27.58 & 139.41 & 34.89 & 30.32 & 32.63 & 154.38\\
0130 p14 & 44.51 & 0.13 & 44.56 & 0.23 & 44.55 & 0.14 & 44.03 & 0.71 & * & *\\
0130 p15 & 39.63 & 32.22 & 39.20 & 87.54 & 35.57 & 78.89 & 39.96 & 45.98 & * & *\\
0130 p16 & 45.58 & 0.22 & 45.05 & 1.22 & 45.81 & 0.18 & 45.57 & 0.29 & * & *\\
0130 p17 & 46.00 & 18.23 & 45.19 & 77.79 & 48.64 & 3.12 & 45.93 & 28.43 & 40.11 & 75.89\\
0133 p01 & -2.38 & 0.28 & -1.91 & 0.63 & -2.42 & 0.58 & -2.32 & 0.20 & * & *\\
0133 p02 & 48.17 & 0.51 & 49.68 & 3071663 & 39.01 & 20.46 & 48.18 & 0.90 & * & *\\
0133 p03 & 126.26 & 4.01 & 111.71 & 34.89 & 70.64 & 1206.68 & 128.23 & 1.32 & * & *\\
0133 p04 & 136.41 & 1.10 & 139.91 & 0.97 & 135.35 & 1.29 & 136.41 & 1.44 & 136.42 & 3.60\\
0148 p01 & -3.13 & 0.39 & -1.93 & 0.32 & -3.39 & 0.61 & -21.97 & 22.71 & * & *\\
0148 p02 & 1.14 & 0.52 & 1.40 & 0.24 & 1.07 & 1.41 & 1.92 & 0.01 & * & *\\
0148 p03 & 6.89 & 1.08 & 5.31 & 2.56 & 7.89 & 1.04 & -11.98 & 111.59 & * & *\\
0160 p01 & -4.93 & 8.42 & -8.43 & 49.92 & -6.54 & 15.96 & -8.33 & 39.27 & * & *\\
0171 p01 & -6.78 & 1.07 & -9.15 & 2.75 & -4.85 & 0.86 & -4.92 & 1.20 & -4.79 & 5177096.8\\
0171 p02 & 14.34 & 0.10 & 14.34 & 13331.5 & 14.34 & 0.10 & 14.34 & 1631526.4 & 14.33 & 2.08\\
0179 p01 & -1.25 & 0.16 & -2.21 & 0.48 & -1.35 & 0.28 & -0.66 & 0.06 & 0.05 & 7.83\\
0179 p02 & -13.21 & 50.78 & -36.83 & 1036.9 & -40.00 & 484.20 & -12.20 & 48.00 & * & *\\
0211 p01 & -3.32 & 0.54 & -3.53 & 1.39 & -2.82 & 0.67 & -3.58 & 0.87 & -2.26 & 0.72\\
0214 p01 & -1.36 & 0.54 & -4.03 & 6.38 & -0.99 & 0.30 & -1.09 & 0.71 & -0.12 & 0.65\\
0214 p02 & 4.90 & 10.95 & 0.00 & 90.05 & -12.55 & 347.75 & 8.40 & 0.80 & 7.61 & 0.70\\
0214 p03 & 7.47 & 4.94 & -4.29 & 130.4 & 9.04 & 2.44 & 9.02 & 2.50 & 9.79 & 4.81\\
0219 p01 & -3.06 & 1.38 & -3.26 & 1.69 & -1.95 & 0.93 & -2.75 & 5.49 & -3.01 & 82.63\\
0219 p02 & 102.97 & 2.35 & 101.29 & 14.94 & 100.46 & 9.93 & 103.06 & 2.41 & 104.17 & 6.01\\
0219 p03 & 126.43 & 0.02 & 126.53 & 1.00 & 126.41 & 0.03 & 126.32 & 0.05 & * & *\\
0219 p04 & 129.98 & 0.18 & 129.94 & 0.41 & 130.27 & 0.13 & 129.97 & 0.27 & * & *\\
0219 p05 & 132.5 & 0.05 & 132.45 & 0.12 & 132.59 & 0.05 & 132.46 & 0.10 & 132.97 & 655.72\\
0222 p01 & -0.90 & 0.10 & -0.86 & 0.18 & -0.70 & 0.11 & -0.97 & 0.15 & * & *\\
0222 p02 & 55.54 & 0.08 & 55.57 & 0.20 & 55.38 & 0.15 & 55.7 & 0.10 & * & *\\
0222 p03 & 61.62 & 0.11 & 53.04 & 11.95 & 61.78 & 0.05 & 61.56 & 0.18 & * & *\\
0222 p04 & 19.22 & 103.14 & 9.58 & 265.33 & 0.60 & 221.82 & 1.75 & 221.29 & * & *\\
0226 p01 & 19.10 & 140.55 & 8.22 & 916.94 & 19.33 & 212.01 & 7.87 & 413.76 & * & *\\
0226 p02 & 37.27 & 19.61 & 47.64 & 0.72 & 39.76 & 13.58 & 38.66 & 95.56 & * & *\\
0226 p03 & 32.00 & 65.12 & 18.55 & 353.98 & 32.44 & 159.72 & 44.07 & 0.48 & * & *\\
0226 p04 & 50.53 & 0.72 & 20.67 & 126.6 & 50.89 & 0.95 & 50.09 & 0.98 & 47.98 & 14.63\\
0226 p05 & -61.40 & 174.64 & -0.77 & 198.95 & -63.35 & 271.36 & -122.68 & 458.26 & -67.40 & 3616.4\\
0226 p06 & 89.53 & 1.04 & 81.34 & 30.63 & 89.49 & 1.61 & 89.00 & 2.02 & * & *\\
0226 p07 & 97.70 & 0.49 & 89.05 & 29.53 & 97.93 & 0.44 & 95.85 & 1.85 & * & *\\
0228 p01 & -0.78 & 0.10 & -0.47 & 0.04 & -1.22 & 0.35 & -0.70 & 0.13 & * & *\\
0237 p01 & -8.35 & 4.80 & -3.78 & 2.69 & -4.22 & 2.29 & -45.03 & 146.18 & -12.61 & 174.95\\
0237 p02 & 0.00 & 24.72 & -12.05 & 662.50 & -9.08 & 724.84 & -11.53 & 788.68 & -0.48 & 776.58\\
0257 p01 & -0.16 & 0.02 & -0.21 & 0.02 & -0.19 & 0.04 & -0.13 & 0.22 & -0.38 & 256027.56\\
0257 p02 & 9.01 & 0.31 & 8.76 & 0.40 & 8.42 & 0.75 & 9.68 & 93238.64 & * & *\\
0269 p01 & -0.44 & 0.03 & -0.42 & 0.09 & -0.44 & 0.07 & -0.59 & 0.06 & -27.88 & 556.91\\
0288 p01 & -3.10 & 0.32 & -5.45 & 0.72 & -3.30 & 0.16 & -3.78 & 0.38 & -3.19 & 1416846.5\\
0288 p02 & 6.80 & 1.00 & 7.95 & 0.70 & 6.15 & 3.84 & 8.12 & 4810513.6 & * & *\\
0332 p01 & -1.30 & 0.09 & -0.49 & 0.14 & -1.60 & 0.18 & -2.40 & 0.22 & * & *\\
0351 p01 & 11.44 & 3.33 & 12.64 & 5.34 & 12.59 & 4.63 & 16.45 & 3.35 & 7.18 & 57.04\\
0351 p02 & 52.71 & 1.26 & 54.67 & 1.62 & 52.63 & 42.97 & 51.34 & 2.74 & 60.06 & 3.23\\
0351 p03 & -27.85 & 7.78 & -10.86 & 2.22 & -7.50 & 1.32 & -26.21 & 9.40 & -3.17 & 2029129.7\\
0398 p01 & -6.86 & 0.67 & -6.92 & 1.25 & -6.98 & 0.96 & -12.53 & 3.13 & -7.71 & 14.66\\
0398 p02 & 7.35 & 0.16 & 7.93 & 0.13 & 6.69 & 0.38 & 7.47 & 0.25 & 7.49 & 435.06\\
0401 p01 & -3.99 & 1.34 & -2.72 & 0.86 & -2.05 & 0.24 & -2.15 & 0.45 & * & *\\
0404 p01 & -6.00 & 0.50 & -6.20 & 1.00 & -3.76 & 0.47 & -4.40 & 0.57 & * & *\\
0404 p02 & 5.27 & 35.67 & -6.76 & 215.17 & -20.15 & 322.64 & -58.67 & 2051.65 & * & *\\
0404 p03 & 31.71 & 0.07 & 31.39 & 0.50 & 31.53 & 0.19 & 31.88 & 0.08 & * & *\\
0404 p04 & 33.70 & 0.05 & 33.36 & 0.32 & 33.72 & 0.08 & 33.77 & 0.03 & * & *\\
0408 p01 & 4.79 & 0.14 & 4.85 & 0.20 & 4.58 & 0.55 & 4.86 & 0.14 & 4.86 & 1.40\\
0408 p02 & 20.78 & 0.05 & 20.78 & 0.29 & 20.80 & 0.07 & 20.80 & 0.06 & * & *\\
0408 p03 & 21.14 & 0.27 & 20.71 & 1.37 & 20.81 & 0.75 & 21.23 & 0.36 & 20.66 & 12.08\\
0408 p04 & 24.66 & 0.07 & 24.91 & 709.32 & 24.59 & 0.18 & 24.53 & 0.28 & 24.55 & 0.88\\
0408 p05 & 27.85 & 2.18 & 27.9 & 7.58 & 22.05 & 25.39 & 28.27 & 2.33 & * & *\\
0408 p06 & 32.79 & 0.16 & 32.38 & 0.64 & 32.61 & 0.34 & 32.68 & 0.23 & * & *\\
0408 p07 & 34.30 & 0.37 & 34.39 & 0.54 & * & * & 34.27 & 0.49 & * & *\\
0408 p08 & 35.58 & 1.21 & 34.78 & 3.43 & 35.52 & 1.73 & 37.19 & 0.27 & 38.31 & 58.27\\
0408 p09 & 41.53 & 1.72 & 41.39 & 6.61 & 42.9 & 0.93 & 42.59 & 3.06 & 43.35 & 1.04\\
0408 p10 & 48.56 & 0.27 & 46.59 & 1.69 & 48.23 & 0.45 & 49.28 & 0.13 & 50.82 & 0.92\\
0408 p11 & 53.37 & 0.12 & 53.44 & $3\times10^{-6}$ & 53.02 & 0.64 & 53.33 & 0.08 & 52.67 & 3560691.4\\
0408 p12 & 59.30 & 2.73 & 59.77 & 7.54 & 58.09 & 5.37 & 60.01 & 2.73 & * & *\\
0408 p13 & 65.94 & 0.19 & 65.57 & 0.81 & 65.97 & 0.29 & 66.27 & 232.52 & * & *\\
0408 p14 & 71.65 & 2.26 & 71.14 & 6.70 & 59.4 & 35.02 & 71.36 & 4.10 & 75.61 & 7.54\\
0408 p15 & 80.70 & 5.66 & 82.21 & 12.43 & 72.22 & 3.43 & 80.57 & 6.91 & 78.96 & 43.36\\
0408 p16 & 92.42 & 4.73 & 92.48 & 13.82 & 81.05 & 69.70 & 92.54 & 6.65 & 91.6 & 27.40\\
0414 p01 & -39.50 & 258.82 & -5.00 & 1.79 & -34.10 & 355.4 & -57.62 & 1212.8 & * & *\\
0451 p01 & -1.93 & 0.83 & -2.23 & 2.73 & -1.08 & 0.59 & -2.03 & 1.06 & -1.48 & 1.48\\
0451 p02 & 0.17 & 0.04 & 0.19 & 0.09 & 0.12 & 0.06 & 0.15 & 0.06 & 0.38 & 108854.65\\
0451 p03 & 3.78 & 0.01 & 3.77 & 0.03 & 3.76 & 0.03 & 3.79 & 0.01 & 3.90 & 0.72\\
0451 p04 & 4.37 & 0.08 & 4.22 & 0.15 & 4.30 & 0.15 & 4.49 & 0.14 & 3.74 & 15.1\\
0451 p05 & 5.31 & 0.03 & 5.36 & 0.09 & 5.32 & 0.04 & 5.17 & 0.12 & 5.41 & 491.81\\
0465 p01 & -3.30 & 0.19 & -3.26 & 0.26 & -3.08 & 0.17 & -3.38 & 0.17 & -2.82 & 3637.69\\
0467 p01 & -0.38 & 0.06 & -0.50 & 0.22 & -0.71 & 0.23 & -0.25 & 0.05 & 0.06 & 0.04\\
0467 p02 & -0.47 & 0.05 & -1.43 & 0.35 & -0.66 & 0.13 & -0.38 & 0.05 & 0.00 & 0.001\\
0469 p01 & 1.35 & 0.13 & 1.22 & 0.55 & 1.85 & 0.08 & 1.29 & 0.17 & 1.07 & 0.42\\
0469 p02 & -0.89 & 0.08 & -0.73 & 0.25 & -1.19 & 0.26 & -0.74 & 0.07 & -0.64 & 0.05\\
0469 p03 & 3.32 & 0.09 & 3.03 & 0.38 & 3.45 & 0.15 & 3.33 & 0.11 & 3.56 & 0.27\\
0472 p01 & -3.77 & 0.39 & -189.60 & 1965.58 & -6.96 & 2.37 & -4.57 & 1.25 & * & *\\
0473 p01 & -1.96 & 0.10 & -2.27 & 0.37 & -1.69 & 0.06 & -1.90 & 0.11 & 1.09 & 679965.67\\
0493 p01 & -1.90 & 0.35 & -3.92 & 1.80 & -1.73 & 0.41 & -1.64 & 0.51 & * & *\\
0501 p01 & -0.87 & 0.05 & -0.95 & 0.14 & -0.85 & 0.08 & -0.98 & 0.09 & * & *\\
0516 p01 & -9.57 & 3.07 & -7.95 & 2.45 & -15.93 & 6.15 & -4.06 & 1.89 & -15.79 & 81.87\\
0516 p02 & -0.13 & 0.05 & -0.13 & $4\times10^{-6}$ & -0.13 & 0.01 & -0.13 & 0.04 & -0.13 & 1.09\\
0537 p01 & -0.24 & 0.02 & -0.13 & 0.04 & -0.13 & 0.04 & -0.43 & 0.06 & -0.76 & 0.62\\
0540 p01 & -0.71 & 0.035 & -0.71 & 0.05 & -0.70 & 0.04 & -0.64 & 0.09 & * & *\\
0543 p01 & 2.94 & 0.04 & 2.79 & 0.08 & 3.11 & 0.04 & 2.88 & 0.06 & 3.17 & 6.51\\
0548 p01 & -0.53 & 0.06 & -0.84 & 0.68 & -0.60 & 0.14 & -0.56 & 0.06 & * & *\\
0548 p02 & -0.51 & 1.78 & -9.53 & 39.03 & -2.58 & 5.46 & -0.35 & 1.78 & * & *\\
0550 p01 & -1.71 & 0.09 & -1.75 & 0.13 & -1.75 & 0.13 & -1.66 & 0.20 & * & *\\
0563 p01 & -0.75 & 0.04 & -0.80 & 0.17 & -1.28 & 0.13 & -1.00 & 0.06 & -0.37 & 0.09\\
0577 p01 & -3.96 & 0.40 & -2.97 & 0.69 & -3.90 & 0.64 & -5.44 & 0.77 & * & *\\
0591 p01 & -13.28 & 3.16 & -6.74 & 0.73 & * & * & -8.96 & 1.34 & * & *\\
0591 p02 & 23.39 & 585509.49 & 23.39 & 844716.06 & 23.42 & 0.26 & 23.30 & 1554776.70 & * & *\\
0591 p03 & 24.47 & 0.18 & 24.42 & 0.55 & 24.31 & 0.32 & 24.71 & 0.39 & * & *\\
0591 p04 & 28.09 & 1.15 & 26.82 & 3.32 & 23.12 & 13.08 & 28.76 & 1.31 & * & *\\
0591 p05 & 30.17 & 0.61 & 31.1 & 0.84 & 30.51 & 0.30 & 29.47 & 0.69 & * & *\\
0591 p06 & 39.81 & 0.28 & 39.81 & 1710836.2 & 39.80 & 0.21 & 39.80 & 0.21 & * & *\\
0593 p01 & -0.82 & 0.16 & -0.96 & 0.64 & -0.98 & 0.53 & -1.36 & 0.71 & * & *\\
0593 p02 & 0.76 & 0.17 & 0.93 & 0.13 & -2.56 & 6.27 & 0.76 & 0.45 & * & *\\
0593 p03 & 4.34 & 0.28 & 4.59 & 0.20 & 4.08 & 0.54 & 4.25 & 0.63 & * & *\\
0594 p01 & -2.28 & 0.17 & -2.95 & 0.90 & -2.22 & 0.22 & -2.05 & 0.14 & * & *\\
0594 p02 & -6.72 & 34.06 & -2.97 & 70.11 & -6.51 & 76.07 & -7.27 & 32.19 & * & *\\
0594 p03 & 4.56 & 0.18 & 4.88 & 985.26 & 4.56 & 24770.6 & 4.46 & 0.17 & * & *\\
0594 p04 & 5.24 & 0.73 & 5.97 & 942.81 & 5.12 & 1.50 & 5.10 & 1.06 & * & *\\
0594 p05 & 7.20 & 0.57 & 6.75 & 2.43 & 7.62 & 0.32 & 7.31 & 0.66 & * & *\\
0594 p06 & 9.15 & 0.54 & 9.65 & 0.23 & 9.08 & 1.63 & 9.30 & 0.26 & * & *\\
0594 p07 & 10.22 & 0.27 & 9.04 & 1.22 & 9.81 & 1.25 & 10.19 & 0.22 & 10.72 & 0.37\\
0594 p08 & 18.79 & 0.36 & 19.01 & 0.20 & 18.5 & 1.63 & 18.72 & 0.34 & * & *\\
0594 p09 & 17.07 & 1131238.30 & 17.15 & $1\times10^{-7}$ & 17.15 & 0.55 & 16.64 & 4628519.40 & * & *\\
0594 p10 & 16.53 & 15.95 & 18.07 & 51.71 & 16.28 & 30.19 & 16.80 & 14.80 & * & *\\
0594 p11 & 23.03 & 0.75 & 23.59 & 0.58 & 22.95 & 1.09 & 23.19 & 1.04 & * & *\\
0606 p01 & -6.55 & 1.85 & -6.37 & 3.35 & -13.02 & 10.13 & -4.21 & 1.15 & * & *\\
0606 p02 & 4.85 & 0.17 & 5.52 & 0.25 & 4.46 & 0.30 & 5.12 & 0.17 & * & *\\
0612 p01 & -33.44 & 13.11 & -13.49 & 6.21 & * & * & -13.00 & 3.48 & * & *\\
0612 p02 & 4.71 & 0.13 & 4.72 & 0.75 & 4.80 & 0.10 & 4.51 & 8.84 & -27.57 & 668.10\\
0612 p03 & 6.58 & 0.08 & 5.07 & 1.99 & 6.62 & 0.15 & 6.35 & 0.23 & 6.16 & 0.69\\
0630 p01 & -17.48 & 4.30 & -11.72 & 4.64 & -23.17 & 10.76 & -30.75 & 23.46 & * & *\\
0658 p01 & -4.20 & 0.20 & -3.31 & 0.28 & -4.10 & 0.30 & -4.00 & 0.27 & * & *\\
0659 p01 & -3.70 & 0.75 & -6.88 & 7.99 & -2.80 & 0.76 & -4.37 & 1.07 & -3.62 & 5.44\\
0659 p02 & -28.75 & 6.10 & -19.47 & 6.04 & -24.39 & 7.82 & -18.30 & 11.06 & * & *\\
0659 p03 & 45.78 & 0.06 & 45.70 & 0.13 & 45.87 & 0.03 & 44.50 & 1.16 & * & *\\
0660 p01 & -0.10 & 0.01 & -0.06 & 0.05 & -0.11 & 0.02 & -0.11 & 0.02 & -1.06 & 0.69\\
0660 p02 & 2.54 & 0.15 & 2.83 & 0.16 & 2.38 & 0.24 & 2.59 & 0.16 & -3.72 & 17.29\\
0673 p01 & -51.73 & 28.9 & -22.25 & 27.03 & -41.96 & 26.51 & -16.63 & 5.13 & * & *\\
0680 p01 & -10.34 & 3.48 & -4.01 & 1.36 & -10.25 & 4.70 & -41.90 & 81.90 & * & *\\
0685 p01 & -0.384 & 0.02 & -0.384 & 0.002 & -0.384 & 0.02 & -0.57 & 0.09 & * & *\\
0686 p01 & 2.22 & 0.42 & 2.52 & 0.35 & 1.49 & 8.99 & 1.75 & 1.61 & * & *\\
0686 p02 & -1.26 & 0.21 & -1.31 & 0.63 & -1.36 & 0.34 & -1.15 & 0.21 & * & *\\
0692 p01 & -0.58 & 0.04 & -0.39 & 0.08 & -0.57 & 0.05 & -0.57 & 0.04 & * & *\\
0692 p02 & 0.39 & 0.45 & -0.72 & 10.09 & -2.88 & 12.66 & 0.52 & 0.21 & * & *\\
0692 p03 & -1.21 & 7.34 & 3.78 & 0.85 & 2.66 & 1.53 & -2.62 & 15.39 & * & *\\
0704 p01 & -0.28 & 0.02 & -0.32 & 0.31 & -0.13 & 0.14 & -0.13 & 0.01 & -0.27 & 0.16\\
0727 p01 & -3.82 & 0.93 & -3.63 & 2.18 & -5.78 & 2.93 & -4.76 & 1.89 & * & *\\
0741 p01 & -7.82 & 1.49 & -327.17 & 5791.39 & -4.06 & 0.64 & -7.67 & 2.08 & * & *\\
0752 p01 & -1.00 & 0.19 & -0.52 & 0.21 & -0.68 & 0.20 & -1.51 & 0.52 & * & *\\
0752 p02 & -35.61 & 685.97 & -2.32 & 14.71 & -55.59 & 3621.16 & -71.59 & 2452.00 & * & *\\
0753 p01 & -3.26 & 0.47 & -21.74 & 21.82 & * & * & -2.24 & 0.25 & * & *\\
0755 p01 & -0.43 & 0.03 & -0.41 & 0.06 & -0.44 & 0.05 & -0.44 & 0.05 & * & *\\
0764 p01 & -7.85 & 0.57 & -2.46 & 0.28 & -4.40 & 0.30 & -10.25 & 1.17 & * & *\\
0803 p01 & 3.23 & 0.20 & 3.29 & 1073.30 & 3.06 & 0.66 & 3.18 & 0.23 & * & *\\
0803 p02 & -0.89 & 0.17 & -0.15 & 0.05 & -0.61 & 0.18 & -2.48 & 0.78 & -0.88 & 0.37\\
0803 p03 & 4.98 & 0.23 & 4.44 & 2.01 & 4.95 & 0.38 & 5.24 & 0.10 & * & *\\
0803 p04 & 4.36 & 0.24 & 4.32 & 0.85 & 4.41 & 0.39 & 4.07 & 0.94 & * & *\\
0815 p01 & -17.00 & 1.33 & -19.85 & 2.89 & -18.37 & 2.89 & -11.94 & 0.62 & * & *\\
0815 p02 & -1.91 & 0.26 & -2.00 & 0.28 & -1.87 & 0.34 & -1.22 & 0.05 & * & *\\
0815 p03 & 1.87 & 0.43 & 1.84 & 0.80 & 1.91 & 0.46 & 0.56 & 2.28 & * & *\\
0824 p01 & -3.65 & 0.43 & -8.34 & 2.08 & -5.87 & 0.88 & -1.53 & 0.16 & * & *\\
0840 p01 & -0.13 & 1223.58 & -0.13 & 70733.02 & -0.09 & 60057.35 & -0.13 & 111619.26 & -0.07 & 0.02\\
0840 p02 & 2.07 & 0.24 & 2.06 & 0.95 & -2.40 & 25.85 & 2.07 & 0.23 & 2.23 & 0.36\\
0840 p03 & 4.16 & 0.16 & 4.16 & 0.67 & 4.16 & 0.04 & 4.16 & 0.22 & 4.06 & 28.66\\
0840 p04 & 5.12 & 0.07 & 5.14 & 0.34 & 4.95 & 0.17 & 5.13 & 0.06 & -19.31 & 1046.60\\
0867 p01 & -0.40 & 0.06 & -0.23 & 0.06 & -0.15 & 0.013 & -0.72 & 0.14 & -0.87 & 0.46\\
0867 p02 & 1.24 & 0.08 & 1.29 & 0.90 & 0.75 & 0.56 & 1.18 & 0.11 & 1.52 & 0.06\\
1008 p01 & -12.13 & 5.74 & -47.81 & 180.78 & -11.43 & 8.40 & -7.03 & 2.69 & * & *\\
1008 p02 & -0.37 & 0.80 & 0.87 & 0.84 & -1.41 & 2.57 & -0.18 & 0.83 & * & *\\
1008 p03 & -0.80 & 9.58 & 6.40 & 0.80 & -10.24 & 104.69 & -9.23 & 45.49 & * & *\\
1008 p04 & 12.79 & 0.17 & 6.52 & 20.82 & 8.09 & 9.51 & 12.81 & 0.15 & * & *\\
1008 p05 & 3.11 & 12.94 & -17.71 & 228.98 & -6.06 & 131.25 & 7.61 & 3.32 & * & *\\
1009 p01 & -3.08 & 0.36 & -3.20 & 0.45 & -3.06 & 0.46 & -10.64 & 8.85 & * & *\\
1009 p02 & 62.08 & 0.38 & 62.65 & 0.65 & 61.82 & 0.73 & 62.67 & 0.19 & * & *\\
1009 p03 & 84.79 & 1.13 & 87.43 & 1.21 & 85.33 & 1.49 & 82.71 & 2.23 & 83.81 & 8.45\\
1009 p04 & 118.94 & 0.48 & 118.58 & 2.12 & 117.73 & 1.01 & 120.24 & 0.24 & 118.16 & 2.83\\
1025 p01 & 0.75 & 0.01 & 0.75 & 0.02 & 0.74 & 0.01 & 0.78 & 0.06 & * & *\\
1025 p02 & -10.67 & 156.87 & -18.43 & 1191.60 & -17.34 & 527.65 & -10.89 & 167.49 & * & *\\
1025 p03 & 1.59 & 0.04 & 1.49 & 0.15 & 1.38 & 0.14 & 1.63 & 0.03 & 1.65 & 0.10\\
1025 p04 & -24.98 & 169.13 & -3.76 & 7.83 & -32.68 & 428.31 & -27.76 & 274.91 & * & *\\
1039 p01 & -2.19 & 0.40 & -2.01 & 0.72 & -2.31 & 0.69 & -1.98 & 0.36 & * & *\\
1039 p02 & -0.73 & 0.82 & -0.04 & 2.46 & -0.63 & 0.94 & -2.03 & 2.12 & * & *\\
1042 p01 & -20.21 & 0.14 & -20.37 & 0.22 & -20.89 & 0.56 & -20.56 & 0.59 & * & *\\
1042 p02 & -0.08 & 1.59 & -0.32 & 3.50 & 1.20 & 0.26 & -2.59 & 5.46 & * & *\\
1042 p03 & -3.01 & 2.89 & -5.35 & 10.01 & -1.34 & 0.36 & -3.07 & 6.12 & * & *\\
1200 p01 & -12.11 & 1.43 & -17.39 & 4.52 & -8.10 & 1.22 & -4.24 & 0.67 & * & *\\
1406 p01 & -1.51 & 0.08 & -1.58 & 0.23 & -1.24 & 0.12 & -1.41 & 0.10 & -0.45 & 0.03\\
1561 p01 & -1.65 & 0.17 & -5.56 & 1.84 & -1.78 & 0.31 & -1.57 & 0.18 & * & *\\
2600 p01 & -2.26 & 0.27 & -6.80 & 3.28 & -2.89 & 0.66 & -173.67 & 1115.52 & -6.01 & 498.93\\
2600 p02 & 6.41 & 0.34 & 6.86 & 0.60 & 6.1 & 0.48 & 4.40 & 0.85 & * & *\\
\enddata
\end{deluxetable}

\clearpage

\begin{deluxetable}{ccccccccccc}
\tabletypesize{\scriptsize}
\rotate
\tablecaption{Start times for some Short GRB pulses. Start times are given in seconds since the trigger, and uncertainties are given in seconds. An asterisk (*) indicates insufficient signal in the energy channel to fit the pulse. \label{tbl-2}}
\tablewidth{0pt}
\tablehead{
\colhead{BATSE Pulse} & \colhead{$t_{s;\rm all}$} & $\sigma_{t_{s;\rm all}}$ & \colhead{$t_{s;1}$} & \colhead{$\sigma_{t_{s;1}}$} &
\colhead{$t_{s;2}$} & \colhead{$\sigma_{t_{s;2}}$} & \colhead{$t_{s;3}$} & \colhead{$\sigma_{t_{s;3}}$} & \colhead{$t_{s;4}$} &
\colhead{$\sigma_{t_{s;4}}$}
}
\startdata
BATSE 0108 p01 & -0.16 & 0.05 & -6.79 & 269.47 & -6.81 & 272.11 & -0.11 & 125624.28 & -0.13 & 162422.00\\
BATSE 0206 p01 & -0.16 & 0.02 & -0.21 & 0.16 & -0.17 & 0.04 & -0.16 & 0.02 & -0.13 & 0.02\\
BATSE 0207 p01 & -0.15 & 3.39 & -0.07 & 5.18 & -0.17 & 1.07 & -0.14 & 4.91 & -0.10 & 92447.97\\
BATSE 0218 p01 & -1.10 & 0.15 & -0.77 & 843327.44 & -2.18 & 1.41 & -1.03 & 0.14 & -15.84 & 626.42\\
BATSE 0229 p01 & -0.13 & 0.01 & -0.69 & 1.53 & -0.13 & 78.74 & -0.12 & 374914.64 & * & *\\
BATSE 0254 p01 & -8.51 & 55.48 & -21.16 & 1284.00 & -22.51 & 922.45 & -15.73 & 193.23 & -1.95 & 7.02\\
BATSE 0297 p01 & -3.54 & 5.72 & -4.94 & 58.55 & -4.01 & 14.88 & -12.92 & 78.95 & * & *\\
BATSE 0297 p02 & -0.77 & 0.01 & -0.77 & 0.08 & -0.77 & 0.20 & -0.77 & 0.02 & -0.15 & 0.09\\
BATSE 0298 p01 & 0.18 & 0.01 & 0.26 & 0.28 & 0.18 & 0.01 & 0.17 & 0.01 & 0.24 & 604.7\\
BATSE 0298 p02 & -0.1280001 & $4\times10^{-6}$ & -0.128001 & $8.4\time10^{-5}$ & -0.1280002 & $3\times10^{-6}$ & -0.1280009 & $1.1\times10^{-5}$ & -0.1280007 & $2.4\times10^{-5}$\\
BATSE 0432 p01 & -0.21 & 319.45 & -0.20 & 513062.96 & -0.19 & 110577.24 & -0.3 & 700.62 & -0.20 & 4445.76\\
BATSE 0444 p01 & -4.61 & 3.74 & -1.38 & 0.55 & -2.90 & 2.33 & -9.70 & 29.56 & -5.94 & 366.79\\
BATSE 0474 p01 & -3.61 & 5.05 & -8.11 & 104.40 & -7.5 & 48.79 & -7.42 & 24.89 & -0.59 & 0.31\\
BATSE 0480 p01 & -0.08 & 27.69 & -0.08 & 104.67 & -0.08 & 11.85 & -0.08 & 1015.95 & -0.08 & 29343.63\\
BATSE 0486 p01 & -10.75 & 116.79 & -2.33 & 11.27 & -22.81 & 845.48 & -22.20 & 1.12 & * & *\\
BATSE 0512 p01 & -0.13 & 0.04 & -3.51 & 600.78 & -0.13 & 167043 & -0.128 & 0.053 & -0.13 & 105.81\\
BATSE 0547 p01 & -12.18 & 71.10 & -1.83 & 4.13 & -31.32 & 807.60 & -17.80 & 179.56 & -13.97 & 718.09\\
BATSE 0551 p01 & -4.68 & 12.13 & -0.41 & 0.25 & -0.55 & 0.26 & -10.09 & 67.02 & -8.36 & 109.42\\
BATSE 0555 p01 & -4.78 & 25.61 & -14.52 & 1096.79 & -0.41 & 0.33 & -0.27 & 0.08 & -11.25 & 431.67\\
BATSE 0568 p01 & -1.24 & 2.16 & -0.55 & 1.27 & -3.25 & 27.78 & -0.45 & 0.38 & -3.19 & 231.76\\
BATSE 0575 p01 & -0.064 & $2\times10^{-6}$ & -0.064 & $6\times10^{-6}$ & -0.064 & $4\times10^{-6}$ & -0.064 & $3\times10^{-6}$ & -0.064 & 119.00\\
BATSE 0575 p02 & 0.064 & 0.048 & 0.1 & 0.14 & -0.07 & 0.19 & 0.06 & 0.07 & 0.1 & 0.45\\
BATSE 0603 p01 & -0.26 & 0.04 & -1 & 3.22 & -0.24 & 0.06 & -0.19 & 0.04 & -0.21 & 1.18\\
BATSE 0603 p02 & -0.06 & 0.34 & 0.23 & 0.05 & -0.03 & 0.38 & 0.26 & 0.03 & * & *\\
BATSE 0603 p03 & 0.62 & 192411.05 & 0.64 & 0.03 & 0.6 & 424750.00 & 0.6 & 0.02 & * & *\\
BATSE 0666 p01 & 1.54 & 0.05 & 1.44 & 0.04 & 1.54 & 0.01 & 1.54 & 0.05 & * & *\\
BATSE 0666 p02 & 2.43 & 0.15 & 2.70 & 0.13 & 2.17 & 0.47 & 2.29 & 0.24 & * & *\\
BATSE 0677 p01 & -0.011 & 252.57 & -0.011 & 98118.75 & -0.011 & 221.4 & -0.011 & 81353.91 & -0.009 & 37848.04\\
BATSE 0729 p01 & -0.16 & 0.04 & -0.18 & 0.69 & -0.13 & 0.06 & -0.24 & 0.10 & -0.30 & 0.68\\
BATSE 0734 p01 & -2.09 & 2.19 & -7.72 & 134.03 & -2.33 & 5.62 & -1.17 & 0.67 & -1.80 & 3.52\\
BATSE 0788 p01 & -0.064 & 0.07 & -0.064 & 97218.13 & -0.064 & 0.07 & -0.064 & $2\times10^{-5}$ & * & *\\
BATSE 0799 p01 & -0.064 & $4\times10^{-5}$ & -0.064 & 0.15 & -0.064 & $4\times10^{-5}$ & -0.064 & $1\times10^{-4}$ & * & *\\
BATSE 0809 p01 & -0.11 & 258 & -0.1 & 58478.4 & -0.17 & 1767.3 & -0.13 & 13.27 & -0.14 & 12265718.00\\
BATSE 0809 p02 & -0.42 & 2.15 & -0.37 & 38.72 & 0.26 & 0.33 & -4.71 & 130.2 & -0.84 & 82.47\\
BATSE 0809 p03 & 0.03 & 3579.66 & -0.07 & 19.38 & 0.03 & 2.01 & -0.48 & 240.75 & 0.03 & 2504487.5\\
BATSE 0830 p01 & -0.13 & 0.05 & -0.19 & 0.28 & -0.16 & 0.09 & -0.19 & 0.11 & -0.06 & 0.11\\
BATSE 0834 p01 & -1.39 & 0.22 & -1.39 & 0.49 & -1.05 & 0.08 & -1.16 & 0.03 & -1.21 & 2820544.7\\
BATSE 0856 p01 & -0.19 & 0.10 & -0.14 & 0.06 & -2.68 & 38.90 & -1.30 & 53.25 & * & *\\
BATSE 1051 p01 & -0.16 & 0.10 & -0.17 & 0.06 & -0.14 & 0.24 & -0.12 & 248.15 & * & *\\
BATSE 1073 p01 & -0.21 & 0.03 & -6.74 & 71.6 & -3.23 & 15.86 & -0.19 & 0.03 & -0.16 & 2.13\\
BATSE 1076 p01 & -0.11 & 0.03 & -0.09 & 0.05 & -0.17 & 0.07 & -0.13 & 0.06 & -0.78 & 24.05\\
\enddata
\end{deluxetable}

\clearpage

\begin{table}
\begin{center}
\caption{Long GRB pulses with $\chi^2_i  > 5$ have start times that are potentially inconsistent with the Pulse Start Conjecture. Careful examination of each pulse determines its disposition: pulses with questionable start times as a result of systematic background effects are removed from consideration (X), while those without noticeable systematic effects are retained (O). The large amount of pulse overlap and non-Poisson background results places significant doubt on the measurements of 26 of these 29 discordant pulses. \label{tbl-3}}
\begin{tabular}{cccl}
\tableline\tableline
Pulse & $\chi^2$ & Disposition & Comments\\
\tableline
BATSE 0130 p06 & 6.4 & X &pulse overlap confusion\\
BATSE 0130 p10 & 8.1 & X &pulse overlap confusion\\
BATSE 0130 p11 & 16.8 & X &overlapping pulses; good s/n\\
BATSE 0179 p01 & 78.8 & X &faint pulse may be composed of many pulses\\
BATSE 0228 p01 & 34.4 & X &channel 1 is noisy and may be composed of two pulses\\
BATSE 0288 p01 & 17.1 & X &pulse has low s/n in channel 1\\
BATSE 0332 p01 & 34.4 & O &long single pulse\\
BATSE 0398 p02 & 11.2 & X &may have multiple components (see channel 3) \\
BATSE 0404 p03 & 6.3 & X &pulse overlap confusion\\
BATSE 0404 p04 & 14.2 & X &pulse overlap confusion\\
BATSE 0408 p08 & 12.7 & X &pulse overlap confusion\\
BATSE 0408 p10 & 46.4 & X &pulse overlap confusion\\
BATSE 0467 p01 & 11.7 & X &may be Short GRB pulse\\
BATSE 0467 p02 & 41.6 & X &may be Short GRB extended emission \\
BATSE 0469 p01 & 12.8 & O &overlapping pulses; good s/n\\
BATSE 0473 p01 & 10.1 & X &faint, noisy pulse; multiple components? \\
BATSE 0537 p01 & 11.8 & X &low s/n at low energies\\
BATSE 0543 p01 & 12.3 & X &pulse overlap confusion\\
BATSE 0612 p03 & 9.4 & X &faint, noisy, complex GRB \\
BATSE 0764 p01 & 84.1 & X &secondary, early component visible in channel 3\\
BATSE 0803 p02 & 178.0 & X &2nd of 4 overlapping pulses is very faint in channel 1\\
BATSE 0803 p03 & 6.6 & X &3rd of 4 overlapping pulses is very faint in channel 1\\
BATSE 0815 p01 & 30.5 & X &background source (Vela X-1); pulse overlap confusion\\
BATSE 0815 p02 & 46.2 & X &background source (Vela X-1); pulse overlap confusion\\
BATSE 0824 p01 & 271.2 & X &long single pulse; odd fit in channel 3\\
BATSE 1009 p04 & 17.4 & X &overlaps with two fitted and one unfit pulse\\
BATSE 1042 p02 & 23.2 & X &overlapping pulse is very faint in channel 3\\
BATSE 1200 p01 & 38.3 & O &single pulse\\
BATSE 1561 p01 & 38.7 & X &may be composed of 2 pulses, noticeable in channel 1\\
\tableline
\end{tabular}
\end{center}
\end{table}

\begin{thebibliography}{}
\bibitem[Amati et al.(2002)]{ama02} Amati, L., et al.\ 2002, \aap, 390, 81
\bibitem[Band et al.(1993)]{ban93} Band, D., et al.\ 1993, \apj, 413, 281
\bibitem[Band(1997)]{ban97} Band, D.~L.\ 1997, \apj, 486, 928
\bibitem[Daigne \& Mochkovitch(1998)]{dai98} Daigne, F., \& Mochkovitch, R.\ 1998, \mnras, 296, 275 
\bibitem[Desai(1981)]{des81} Desai, U.~D.\ 1981, \apss, 75, 15 
\bibitem[Hakkila et al.(2000)]{hak00} Hakkila, J., Haglin, D.~J., Pendleton, G.~N., Mallozzi, R.~S., Meegan, C.~A., \& Roiger, R.~J.\ 2000, \apj, 538, 165 
\bibitem[Hakkila et al.(2003)]{hak03} Hakkila, J., Giblin, T.~W., Roiger, R.~J., Haglin, D.~J., 
Paciesas, W.~S., \& Meegan, C.~A.\ 2003, \apj, 582, 320 
\bibitem[Hakkila et al.(2007)]{hak07} Hakkila, J., et al.\ 2007, \apjs, 169, 62
\bibitem[Hakkila et al.(2008)]{hak08} Hakkila, J., et al. \ 2008a, ApJ, 677, L81
\bibitem[Hakkila \& Cumbee(2009a)]{hak09} Hakkila, J., \& Cumbee, R.~S.\ 2009, in AIP Proc. 1133 (ed. Meegan, Gehrels, \& Kouveliotou), 379
\bibitem[Hakkila et al.(2009b)]{hak09b} Hakkila, J., Fragile, P.~C., \& Giblin, T.~W.\ 2009, in AIP Proc. 1133 (ed. Meegan, Gehrels, \& Kouveliotou), 479
\bibitem[Horv{\'a}th(1998)]{hor98} Horv{\'a}th, I.\ 1998, \apj, 508, 757 
\bibitem[Kaneko et al.(2006)]{kan06} Kaneko, Y., Preece, R.~D., Briggs, M.~S., Paciesas, W.~S., Meegan, C.~A., \& Band, D.~L.\ 2006, \apjs, 166, 298 
\bibitem[Katz(1994)]{kat94} Katz, J.~I.\ 1994, \apjl, 432, L107
\bibitem[Kocevski et al.(2003)]{koc03} Kocevski, D., Ryde, F., \& Liang, E.\ 2003, \apj, 596, 389 
\bibitem[Kouveliotou et al.(1993)]{kou93} Kouveliotou, C., Meegan, C.~A., Fishman, G.~J., Bhat, N.~P., Briggs, M.~S., Koshut, T.~M., Paciesas, W.~S., \& Pendleton, G.~N.\ 1993, \apjl, 413, L101 
\bibitem[Liang et al.(1997)]{lia97} Liang, E., Kusunose, M., Smith, I.~A., \& Crider, A.\ 1997, \apjl, 479, L35
\bibitem[Markwardt, C. B.(2008)]{mar08} Markwardt, C.~B., \ 2008, arXiv:0902.2850v1
\bibitem[Mukherjee et al.(1998)]{muk98} Mukherjee~S., et al. 1998, \apj, 508, 314
\bibitem[Nakar \& Piran(2002)]{nak02} Nakar, E., \& Piran, T.\ 2002, \mnras, 331, 40 
\bibitem[Nemiroff et al.(1994)]{nem94} Nemiroff, R.~J., Norris, J.~P., Kouveliotou, C., Fishman, G.~J., Meegan, C.~A., \& Paciesas, W.~S.\ 1994, \apj, 423, 432 
\bibitem[Nemiroff(2000)]{nem00} Nemiroff, R.~J.\ 2000, \apj, 544, 805 
\bibitem[Norris et al.(1996)]{nor96} Norris, J.~P., Nemiroff, R.~J., Bonnell, J.~T., Scargle, 
J.~D., Kouveliotou, C., Paciesas, W.~S., Meegan, C.~A., \& Fishman, G.~J.\ 1996, \apj, 459, 393 
\bibitem[Norris et al.(2000)]{nor00} Norris, J.~P., Marani, G.~F., \& Bonnell, J.~T.\ 2000, \apj, 534, 248 
\bibitem[Norris(2002)]{nor02} Norris, J.~P.\ 2002, \apj, 579, 386 
\bibitem[Norris et al.(2005)]{nor05} Norris, J.~P., Bonnell, J.~T., Kazanas, D., Scargle, J.~D., Hakkila, J., \& Giblin, T.~W.\ 2005, \apj, 627, 324 
\bibitem[Norris \& Bonnell(2006)]{nor06} Norris, J.~P., \& Bonnell, J.~T.\ 2006, \apj, 643, 266 
\bibitem[Paciesas et al.(2000)]{pac99} Paciesas, W.~S., et al.\ 2000, VizieR Online Data Catalog, 9020, 0 
\bibitem[Pendleton et al.(1999)]{pen99} Pendleton, G.~N., et al.\ 1999, \apj, 512, 362 
\bibitem[Peng et al.(2009)]{pen09} Peng, Z.~Y., Ma, L., Zhao, X.~H., Yin, Y., Fang, L.~M., \& Bao, Y.~Y.\ 2009, arXiv:0903.3457 
\bibitem[Piran(2005)]{pir05} Piran, T.\ 2005, Reviews of Modern Physics, 76, 1143 
\bibitem[Ramirez-Ruiz \& Fenimore(2000)]{rrm00} Ramirez-Ruiz, E., \& Fenimore, E.~E.\ 2000, \apj, 539, 712
\bibitem[Reichart et al.(2001)]{rei01} Reichart, D.~E., Lamb, D.~Q., Fenimore, E.~E., Ramirez-Ruiz, E., Cline, T.~L., \& Hurley, K.\ 2001, \apj, 552, 57 
\bibitem[Ryde(2005)]{ryd05a} Ryde, F.\ 2005, \aap, 429, 869
\bibitem[Scargle(1998)]{sca98} Scargle, J.~D.\ 1998, \apj, 504, 405
\bibitem[Schaefer(2007)]{sch07} Schaefer, B.~E.\ 2007, \apj, 660, 16  
\bibitem[Sumner \& Fenimore(1997)]{sum97} Sumner, M.~C., \& Fenimore, E.~E.\ 1997, arXiv:astro-ph/9712302 
\bibitem[Stern \& Svensson(1996)]{ste96} Stern, B.~E., \& Svensson, R.\ 1996, \apjl, 469, L109 
\end{thebibliography}
\end{document}